**Properties of promising cartilage implants based on cellulose/polyacrylamide composite hydrogels: results of *in vivo* tests carried out over a period of 90 – 120 days**

Alexander L. Buyanov[1*], Iosif V. Gofman[1], Svetlana A. Bozhkova[2], Natalia N. Saprykina[1], Georgii I. Netyl'ko[2], Evgenii F. Panarin[1]

[1] Institute of Macromolecular Compounds, Russian Academy of Sciences, St. Petersburg, Russia;

[2] Vreden National Medical Research Center of Traumatology and Orthopedics, St.-Petersburg, Russia

High-strength composite hydrogels "cellulose-polyacrylamide" were synthesized by free-radical polymerization of acrylamide conducted inside the previously formed physical network of regenerated plant cellulose. Partial hydrolysis of the amide groups of these hydrogels yielded their ionic forms with a degree of hydrolysis of 0.1 and 0.25. The cylindrical hydrogel samples of three compositions were implanted in the preformed osteochondral defects of the rabbit's femoral knee joints. No signs of migration or disintegration of the tested implants were revealed in the course of *in vivo* tests as long as 90 and 120 days after the implantation. The mechanical behavior of both the virgin hydrogels-implants and the implants extracted from the joints after *in vivo* experiments was studied in detail. The morphology and chemical composition of the extracted implants were studied by SEM combined with the EDX method. The results obtained were shown that the mechanical characteristics of hydrogel implants remained practically unchanged after *in vivo* tests. The extracted implants, as well as the initial hydrogels, endured cyclic compression loading at the amplitude up to 50 %. Compression stresses up to 3 – 10 MPa were recorded in these tests, which is close to the data obtained by several authors for natural articular cartilages in the same conditions of loading. The principal differences in the chemical composition and morphology of the implant area adjacent to the subchondral bone for non-ionic and ionic types of implants have been revealed. If for non-ionic types of implants in this area intensive mineralization with formation of calcium phosphates inside the polymeric hydrogel network is observed, the border area of ionic implants practically does not undergo mineralization.

* Corresponding authors, E-mail: buyanov799@gmail.com (A.L.B.)

## Introduction

Hydrogels are elastic networks based on hydrophilic polymers swollen in water. Due to high content of water, different types of natural and synthetic hydrogels demonstrate high biocompatibility. Besides, physical properties of these materials resemble properties of living tissues; therefore, they are used as biomaterials for various applications, such as soft wound dressings, contact lenses, drug delivery systems, etc. [1, 2]. Hydrogels have also been the subject of extensive research as replacement materials for damaged cartilage tissue [3 – 11]. In the experiments involving animals, replacement of cartilage tissues of various localizations with hydrogels is studied, including replacement of knee and hip articular cartilages [3, 6 – 8]. "Cartiva", a polyvinyl alcohol-based hydrogel material, has been applied in clinical practice for the replacement of damaged knee joint cartilage [12]. Currently, this material is successfully used for the treatment of arthritis of the finger joint [13].

Despite the progress achieved in this field, there are certain problems that limit the application of hydrogel materials as artificial cartilage tissues, and one of the main problems is the insufficiently high level of their mechanical characteristics. Human articular cartilage is capable of withstanding significant and prolonged mechanical loads, in particular, compressive forces [14]. Healthy joints under conditions of mild to moderate motor activity are subjected to loads resulting in contact stresses of 1 to 6 MPa, but this value can get as high as 12 MPa for more intense activities [15]. It has been estimated that under the so-called 'physiological loading' conditions (about 6 MPa), the value of cartilage compressive deformation varies between 10 and 33% [16]. However, at the maximum loads that can act on cartilage, the magnitude of compression can be as high as 50% [17].

The mechanical characteristics of cartilage samples extracted from joints vary considerably depending on the peculiarities of measurement technique. Under unconfined (with a free lateral surface of a sample) compression, the stress values for cartilage of the dog knee joint (medial femoral condyle, lateral femoral condyle) were ~1.5 and ~6 MPa at deformations of 30% and 50%, respectively [18] in experiments with a deformation rate of 300%/min, which is close to the one used in our work. The compressive strength value for juvenile bovine articular cartilage explants was determined to be loading rate-dependent, reaching a maximum strength of 29.5 ± 4.8 MPa at a deformation rate of 0.10 %/sec, with compressive deformation values as high as 60 - 80 % [19].

For the best types of mono-component hydrogels, including the PVA-based ones, the degree of stress ($\sigma$) under compression of 30–50 % (which corresponds to the "working" strain range of cartilage tissues) does not exceed 1-2 MPa [4, 5, 20].

The hydrogels with improved mechanical properties can be obtained while using cellulose as a reinforcing component in the compositions with other polymers [9]. Cellulose is a natural polymer characterized by high mechanical stiffness and excellent biocompatibility [9, 21], but the major problem in preparing cellulose hydrogels is a lack of appropriate solvents due to its highly extended hydrogen-bonded structure [22] . For this reason, only few examples of the successful formation of cellulose hydrogels are known. One of them was presented in [23]: the irreversible physically crosslinked cellulose gels were prepared from a NaOH–urea aqueous solution system. The hydrogels of this type based on the regenerated plant cellulose withstand the one-shot compression of more than 50 %, the σ value corresponding to this deformation being as high as 0.3 MPa. This level of mechanical properties presumably satisfies the requirements to materials for drug delivery systems and to some other types of biomaterials but is indeed insufficient to use these hydrogels as artificial joint cartilages.

Simple mechanical blending of cellulose with various natural or synthetic polymers can result in an increased level of mechanical performance of composite hydrogels, but the resulting level remains well below that required for cartilage implants. For example, just some little amounts of micro-granulated cellulose being introduced into the polyacrylamide (PAAm) hydrogels at the stage of their synthesis (10 - 150 mg per 1 g of acrylamide) ensure the increase in the Young's modulus of the material up to 1.5 times [24]. However, the absolute value of elastic modulus did not exceed 0.1 MPa, which is an order of magnitude or more lower than the level required for cartilage implants. Besides, many authors have tested different variants of introducing nanocellulose as a reinforcing component into composite hydrogels, both in the form of nanocrystals and nanofibrils, including their modified forms [25 – 28]. In this way, promising types of hydrogels with a useful set of functional characteristics, such as the level of sorption capacity and porosity, were obtained. However, mechanical characteristics of different types of hydrogels reinforced by nano-cellulose are rather modest: the typical values of the stress corresponding to the compression of 30–50% do not exceed several tens of kPa [25, 26, 29 – 31].

The most substantial improvement of the mechanical properties of hydrogels can be obtained in the composite materials with the structure of interpenetrating polymer networks (IPN). In our works, the possibility of synthesizing high-strength composite hydrogels has been shown; these hydrogels were prepared through the formation of IPN in which rigid-chain cellulose of plant or bacterial origin (PC and BC, respectively) serves as a reinforcing "framework", and a synthetic flexible-chain polymer creates the necessary level of swelling and imparts elasticity to the composition. Polyacrylamide (PAAm) was introduced into IPN as a synthetic flexible-chain polymer by radical polymerization of acrylamide monomer. The synthetic method developed in our works allows one to obtain both hydrogel-based highly selective membranes for the separation of aqueous-organic mixtures [32 – 33] and hydrogels for cartilage tissue replacement [34 – 36].

It has been shown that these hydrogels are close to articular cartilages of various types and localizations in terms of mechanical characteristics and viscoelastic behavior. For these gels, as well as for cartilages, the level of stress values fixed in the range of compression deformations from 30 to 50% reaches 5-10 MPa, and they are able to sustainably (i.e., without noticeable deterioration of mechanical characteristics) withstand fatigue tests in the repeated cyclic compression regimes up to deformations of 50-70% [35, 36].

In the series of *in vivo* experiments involving laboratory outbred rats and Chinchilla rabbits, tissue reactions to the introduction of composite hydrogel samples with the IPN structure (hereinafter IPN hydrogels) into soft tissues (muscle), joint cavities, as well as into deep defects of articular cartilage and subchondral bone, were studied at observation periods of up to 90 days [37]. Neither migration nor degradation of the tested samples were detected in these experiments. In none of the experiments performed degenerative or necrotic changes in the tissues surrounding the implants were observed; they did not cause perifocal inflammation, which confirmed their good biocompatibility with muscle, cartilage and bone tissues. It was shown that when cylindrical samples of hydrogels were implanted into the area of a deep cartilage defect of the rabbit knee joint, they functioned effectively as cartilage prosthesis during 90 days of the experiment, ensuring complete preservation of the function of the operated limb.

The present study was aimed at obtaining detailed information on the mechanical behavior of both the initial hydrogel samples to be used as intact implants and the implants extracted from the joints of laboratory animals (Chinchilla rabbits) after their functioning as artificial cartilage for periods of time as high as 90 and 120 days both in the joint areas, which are not subjected to significant mechanical loads and in their “loaded” areas.

It was also interesting to study the structure and morphology of the extracted implants by scanning electron microscopy (SEM), and to investigate their elemental composition by energy-dispersive X-ray spectroscopy (EDX).

Ionic carboxylate groups were introduced into some of the samples to increase the degree of hydrophilicity of implanted hydrogel samples and to control the mineralization process observed earlier in the *in vivo* experiments involving non-ionic hydrogels [38].

## Experimental

### Materials

Acrylamide purchased from Sigma-Aldrich Corp. was recrystallized twice from benzene. All other chemicals of analytical grade were used as received.

To prepare the PC matrices the plant cellulose was dissolved in trifluoroacetic acid during 7 days. The solutions were cast onto the flat glass substrates and dried [33].

The cellulose layers obtained by this way were washed in distilled water up to the neutral pH value of the used fluid. The PC sheets of the thickness of about 5-6 mm were prepared under this protocol. As a starting PC, the cotton linter pulp (cotton cellulose) was used containing ≥ 98.5 wt. % of α-cellulose, less than 0.2 % of ash; the total amount of impurities insoluble in sulfuric acid – less than 0.3 % (purchased at "MegPromTex" Co, Uzbekistan). The degree of polymerization of cotton cellulose was 2500.

The composite hydrogels were synthesized using the technique that we have developed previously, namely by radical copolymerization of acrylamide with the low molecular weight crosslinking agent, N,N'-methylene-bis-acrylamide (MBA) conducted inside the cellulose matrices that were previously swollen in the reaction solution for 24 h [32, 35, 36].

The initial concentrations of acrylamide, MBA and cobalt (III) acetate, used as the initiator, in the water solution were 7.4, $1.4\times10^{-3}$ и $0.5\times10^{-3}$ mol/l, respectively. The extent of the monomer conversion close to 100 % was obtained after 2 h of the reaction at 25 $^{O}$C. Then the composite hydrogels were placed in distilled water for several days to remove the residual amounts of low molecular weight components and to let the gels swell.

To prepare samples containing carboxylate groups, the initial PC-PAAm hydrogels were hydrolyzed in 0.1 M NaOH solution at 25$^{O}$C for 3 and 24 for the samples No 2 and No 4, respectively. The degree of hydrolysis of the amide groups for these samples determined via elemental analysis was 0.10 (No 2) and 0.25 (No 4). Both the swelling of these hydrogels and their testing were carried out in a phosphate-salt buffer solution with pH 7.4.

The hydrogel samples were prepared in the form of the flat plates.

The water content in the samples in the equilibrium state was determined by drying of thin hydrogel plates at 160 $^{O}$C up to the constant weight value. Quantitative chemical composition of the samples was précised both by the elemental analysis and by the gravimetric method basing on the amounts of the components introduced in the reaction.

## Hydrogels characterization

## Mechanical tests

A set of protocols of the mechanical tests has been developed in our previous works to characterize the peculiarities of the mechanical behavior of these materials in the loading modes close to the working conditions of joint cartilage, namely both single-shot compression tests up to the compression value of 80 % or up to the sample's break if the failure occurred below this strain (Fig. 4), and different types of cyclic compression tests (Fig. 5, 6), described in details elsewhere [35, 36]. The cylindrically shaped samples of swollen hydrogel were used in all mechanical tests carried out in this work. These samples of 4-6 mm height and 5-10 mm diameter were cut from the hydrogels' plates in the direction perpendicular to the surface of the plate.

During the single compression tests the stress-strain curves were registered, namely the dependences of the compressive stress $\sigma$ upon the deformation $\varepsilon$ of the samples. To characterize the stiffness of the hydrogels under tests the mean slope value of the compression curves $\Delta\sigma/\Delta\varepsilon$ was determined in the range of the deformation: from 10 to 15% ($E_{10–15\%}$). The stress values were additionally determined corresponding to the compression values of 30 and 50% ($\sigma_{30\%}$ and $\sigma_{50\%}$, respectively).

The most widely used mode of cyclic tests is the multiple cyclic compressions of the hydrogel samples with constant amplitude of the deformation as high as 30-70 %. The tests of the hydrogels under study in these conditions are most suitable to clarify the extent of the stability of the behavior of these materials under the action of the long term mechanical loads with a periodically varying stress.

At these tests the samples of swollen hydrogels are subjected to the multiple unconfined compression acts with a constant deformation amplitude (30-70 %) followed by the decompression with the same speed up to the initial height of the sample. Depending upon the goals of each test its duration can vary in a broad range of cycles' number – from several cycles up to several thousands of cycles. In the tests conducted in this work, the specimens were subjected to 100 cycles with each amplitude of compression. This was followed by relaxation of the specimen in water in the unloaded state, and after that it was exposed to the following series of cycles with increased amplitude (Fig. 2, 3). The maximum stress values corresponding to the amplitude compression values in different cycles were determined in these tests.

**SEM and EDX investigations**

The morphology of the hydrogel samples was studied by SEM method with a Supra 55VP scanning electron microscope (Zeiss, Oberkochen, Germany). Sample chips were made by freezing in liquid nitrogen. The test objects were placed on the objective table and fixed with a conductive scotch tape. To ensure conducting properties of the samples surface and eliminate interference caused by accumulation of the surface charge in the course of scanning, and also to improve the contrast, the sample surface was preliminarily coated with platinum by cathode sputtering with a Quorum 150 installation (Quorum Technologies Ltd, Laughton, the United Kingdom). The coating thickness was about 10 nm. The chips morphology was studied using the secondary electron mode (SE2). The elemental composition of the samples (both of spall regions or on the sample surface) was determined with an INCA Energy-dispersive X-ray microanalysis system (EDX) equipped with an X-Max 80 detector (Oxford Instruments, Abingdon, the United Kingdom) and supplied in combination with the Supra 55VP microscope.

**Biomedical experiments**

For implantation of the hydrogel materials into osteochondral defects of the knee joint cartilage of 18 rabbits of Chinchilla breed cylindrical hydrogel samples were used; their general view is presented in Figure 1.

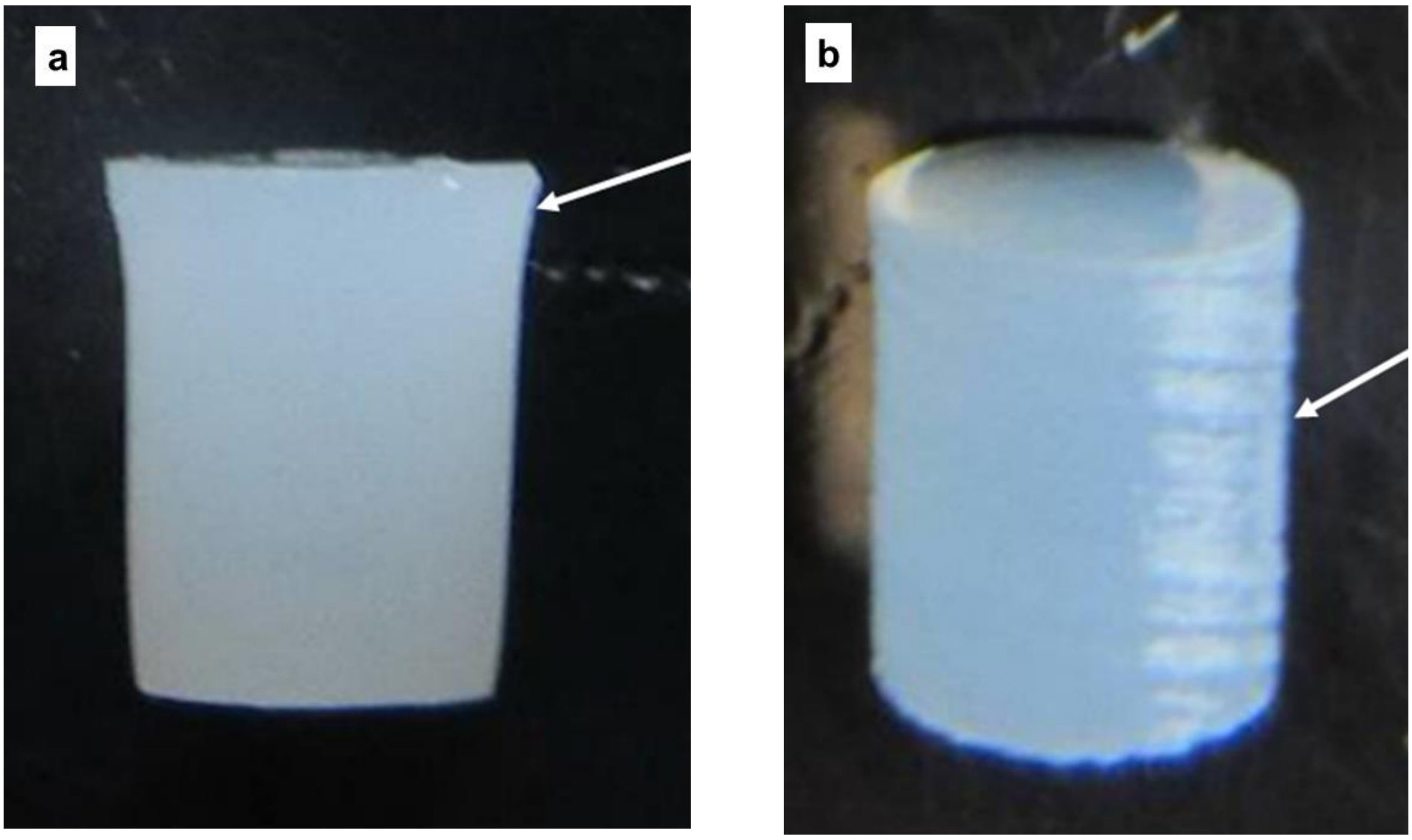

Figure 1. Visual appearance of artificial cartilage samples prepared for implantation.
a – sample 1, b – sample 4. The arrows show the taper and helical rifling for samples 1a and 1b, respectively.

Nonionic sample 1 (Fig. 1a) as well as ionic hydrogel 2 prepared on the basis of hydrogel 1 had a cone-shape for better fixation in the formed cylindrical defects of cartilage and bone tissue. The part of the implant with a larger diameter was placed in the lower part of a defect.

Sample 4 did not have a taper, but a screw thread was cut on its lateral surface; this was also designed to achieve better fixation in the surrounding cartilage and bone tissues (Fig. 1b).

The diameter of cone-shaped samples 1 and 2 was equal to approximately 2.9 and 3.2 mm (the smaller and larger diameters, respectively); the diameter of sample 4 was equal to approximately 3 mm. The heights of all three samples were in the range of 3.7 to 3.9 mm.

The care and use of laboratory animals were performed in accordance with the rules adopted by the European Convention for the Protection of Vertebrate Animals Used for Experimental and Other Purposes, as well as the order of the Ministry of Health of the Russian Federation No. 708N dated August 23, 2010.

All procedures performed on animals have been examineded and approved by the Local Animal Use Ethical Committee for ethical compliance.

All manipulations were performed in a clean operating room of the vivarium of the Russian Research Institute of Traumatology and Orthopaedics named after R.R. Vreden with observance of aseptic and antiseptic rules under intravenous anesthesia (ketamine, diazepam). The prophylaxis of infectious complications was carried out using ceftriaxone.

The hydrogel samples of the three types (1, 2 and 4, see Table 1) were implanted in the preformed osteochondral defects of intercondylar area of the femoral knee joints that are not subjected to significant mechanical loads and of inner femoral condyle ( "loaded" areas), see Figure 2 a and b, respectively.

No cases of implant migration and signs of inflammation were detected at the autopsy of knee joints. In all cases there was a tight adherence of the edges of the tested samples to the edges of the defect (Figure 3) both in the intercondylar area (a) and the area of the inner condyle of the knee joint (b).

The implants were extracted from the joints by cutting out a rectangular area with a cross-section approximately 1 mm larger than their diameter; then they were freed from cartilage and bone tissues with a scalpel. In some cases, the implants were excised together with the adjacent tissues to study their morphology by SEM and EDX.

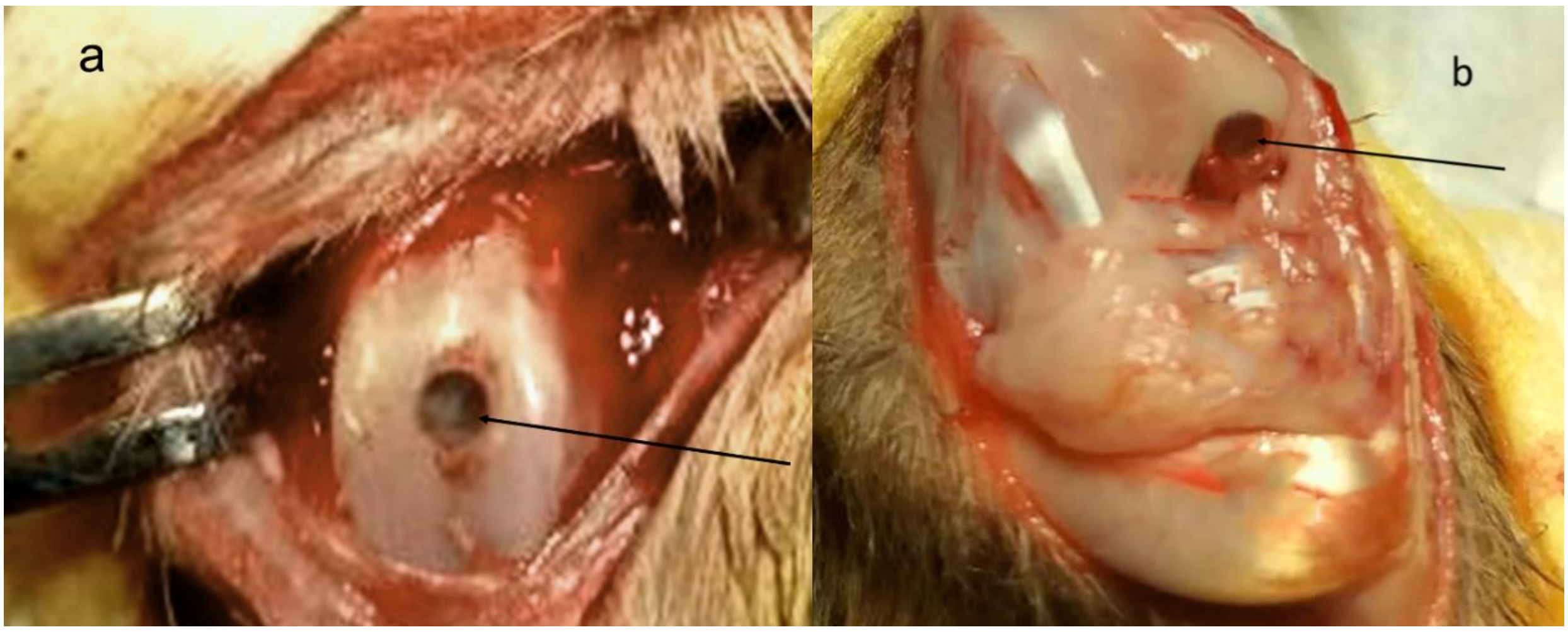

Figure 2. Artificial defects in the intercondylar area (a) and inner femoral condyle (b), designated by arrows.

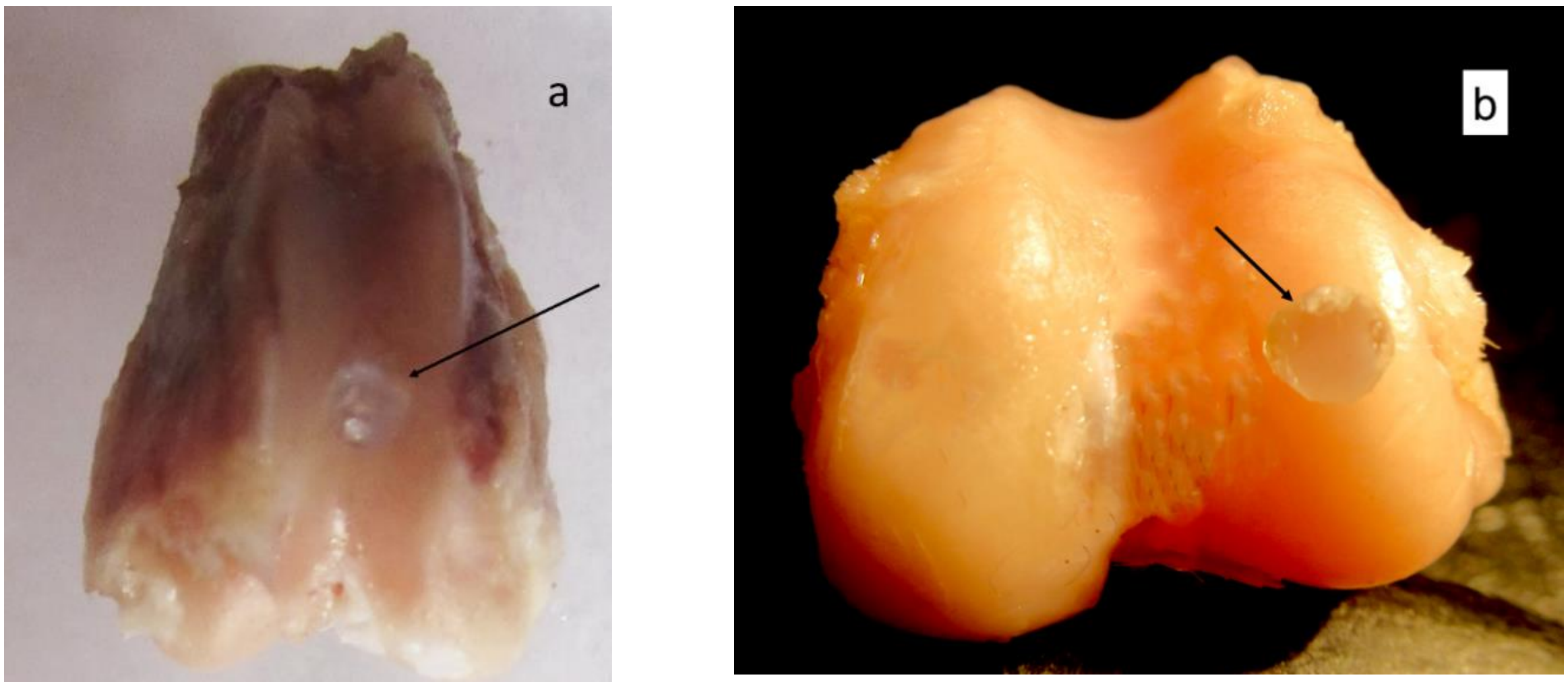

Figure 3. Macropreparations of the rabbit knee joint extracted 90 days after implantation of artificial cartilage samples 1 (a) and 4 (b). The implants are indicated by arrows.

**Results and Discussion**

To investigate the effectiveness of composite hydrogels as cartilage implants, we used the samples with the compositions and characteristics listed in Table 1. The selection of samples was based on the data obtained previously in the study of the synthesis and mechanical behavior of composite hydrogels when varying both the cellulose content and ionic groups content in synthetic polymer chains [35, 36]. The data on biocompatibility of these materials with cartilage and bone tissues of laboratory animals and their ability to maintain integration with these tissues during their functioning as artificial cartilage tissue substitutes were taken into account [37]. The available information about mechanical behavior of natural articular cartilage was also considered.

Table 1. Composition and characteristics of the investigated hydrogels

| Sample No | Sample type | C cell, wt.% | C water, wt.% | E\|10-15%, MPa | σ\|30%, MPa | σ\|50%, MPa |
|---|---|---|---|---|---|---|
| 1 | PC-PAAm | 7 | 63 | 8.14 | 2.71 | 5.24 |
| 2 | PC-PAAm - PAA($Na^+$)_10 | 7 | 72 | 5.92 | 1.68 | 4.50 |
| 3 | PC-PAAm | 18 | 43 | 34.9 | 8.41 | 16.6 |
| 4 | PC-PAAm - PAA($Na^+$)_25 | 18 | 68 | 22.0 | 4.80 | 8.46 |

In this respect, the previously tested samples of hydrogels with cellulose content in the compositions of about 7 and 18 wt.% were of interest. Varying the cellulose content in the compositions enabled us to obtain samples with different equilibrium water contents and different levels of mechanical characteristics. Two non-ionic hydrogels (samples 1 and 3) with cellulose percentages equal to 7 and 18 wt.%, respectively, were used as starting materials for preparation of ionic hydrogels 2 and 4 by alkaline hydrolysis of part of amide groups in PAAm chains. The degrees of hydrolysis of PAAm amide groups were equal to 10 and 25 % for hydrogels 2 and 4, respectively, that is, they contained 10 and 25 mol % of carboxylate groups PAA(Na+) in polymer chains PAAm-PAA(Na+).

The presence of ionic groups in hydrogels leads to an increase in their degrees of swelling for well-known reasons [39], but the mechanical characteristics of hydrogels deteriorate. It is seen from the data presented in Table 1 that for sample 2 (which contains 10 mol % carboxylate groups in synthetic chains polyacrylamide-sodium polyacrylate (PAAm-PAA(Na+)), the equilibrium water content increased by 9 wt.% as compared to that in the non-ionic hydrogel (sample 1), and the level of mechanical properties decreased only slightly. It should be noted that the method of alkaline hydrolysis of PAAm amide groups used in our work to introduce ionic groups into hydrogels does not affect hydrogel structure.

Another sample of non-ionic hydrogel 3, which contains higher amount of cellulose, has too high level of stiffness and at the same time insufficiently high water content (43 wt. %), which is much lower than the range typical of natural cartilages (60 - 80 %) [40]. At the same time, it is seen from Table 1 that ionic hydrogel 4 obtained from non-ionic sample 3 contains more water (by 25 %, i.e., 68 wt.%) and is obviously more suitable as a cartilage substitute in terms of this parameter.

The introduction of ionic groups into hydrogels was also interesting, since it was expected that the presence of these groups would regulate the process of hydrogel mineralization (which has been observed earlier in the course of *in vivo* experiments with non-ionic hydrogels) [38]. The non-ionic hydrogel samples implanted in the area of contact with subchondral bone were subjected to intensive mineralization with the formation of significant amounts of calcium phosphate (up to 40 wt.%) inside the hydrogel, mainly in the form of hydroxyapatite. At the same time, no calcium phosphates were observed in the cartilaginous area of the implants (out of contact with subchondral bone) after 45 days [38]. It should be clarified that in our works, we simulate the possibility of treatment of deep local defects of articular cartilage tissue, and the majority of implants are located in the area of subchondral bone.

On the other hand, the currently available data do not allow one to predict the consequences of long-term exposure of implants to human body; thus, it is necessary to be able to regulate the rate of implant mineralization in order to minimize the possibility of the formation of mineral phase in the cartilage area of implants. Otherwise, the mineralization process can eventually lead to the formation of a bone-like composite in this area, which will not be able to perform the functions of a cartilage tissue substitute.

While the process of implant mineralization in the subchondral region has a positive role, promoting integration between the implant and bone tissue, the mineralization of the cartilaginous region should be avoided for the reasons mentioned above. According to our preliminary data obtained during the *in vitro* study of composite hydrogels mineralization (to be published), the presence of ionic carboxyl groups in these hydrogels reduces the mineralization rate by at least an order of magnitude. In order to investigate the possibility of regulating mineralization rate *in vivo*, the minimum required number of ionic groups was introduced into hydrogels so that their mechanical characteristics remained at a sufficiently high level.

## Mechanical behavior of composite hydrogels

As was mentioned in the Introduction, the level of mechanical characteristics of hydrogels is a key parameter that determines the possibility of their application as artificial cartilage substitutes, especially if they are intended to replace damaged areas of articular cartilages. In this connection, let us now consider mechanical behavior of the hydrogel samples listed in Table 1, which were used in the present work as cartilage implants.

The single compression curves of the hydrogels selected for the experiments are given in Fig. 4. The values of secant moduli of compression obtained in the 10 – 15 % region ($E|_{10-15\%}$) and the stresses corresponding to compressions of 30 and 50 % ($\sigma|_{30\%}$ and $\sigma|_{50\%}$) for these hydrogels are presented in Table 1.

As has been demonstrated in our previous work [35], the single compression curves of hydrogel samples provide information about the initial stiffness (modular characteristics) of hydrogel materials. However, during repeated compression cycles, significant changes in these curves are observed due to the profound reorganization of the system of physical entanglements and hydrogen bonds that existed in the material before the onset of deformation. These changes are reversible, i.e., after a certain period of time (typically more than two days) the shape of the curves is restored to the original one. However, when a certain level of load is exceeded, chemical bonds and cross-linking nodes start to break down, which eventually leads to disintegration of the material [35, 36].

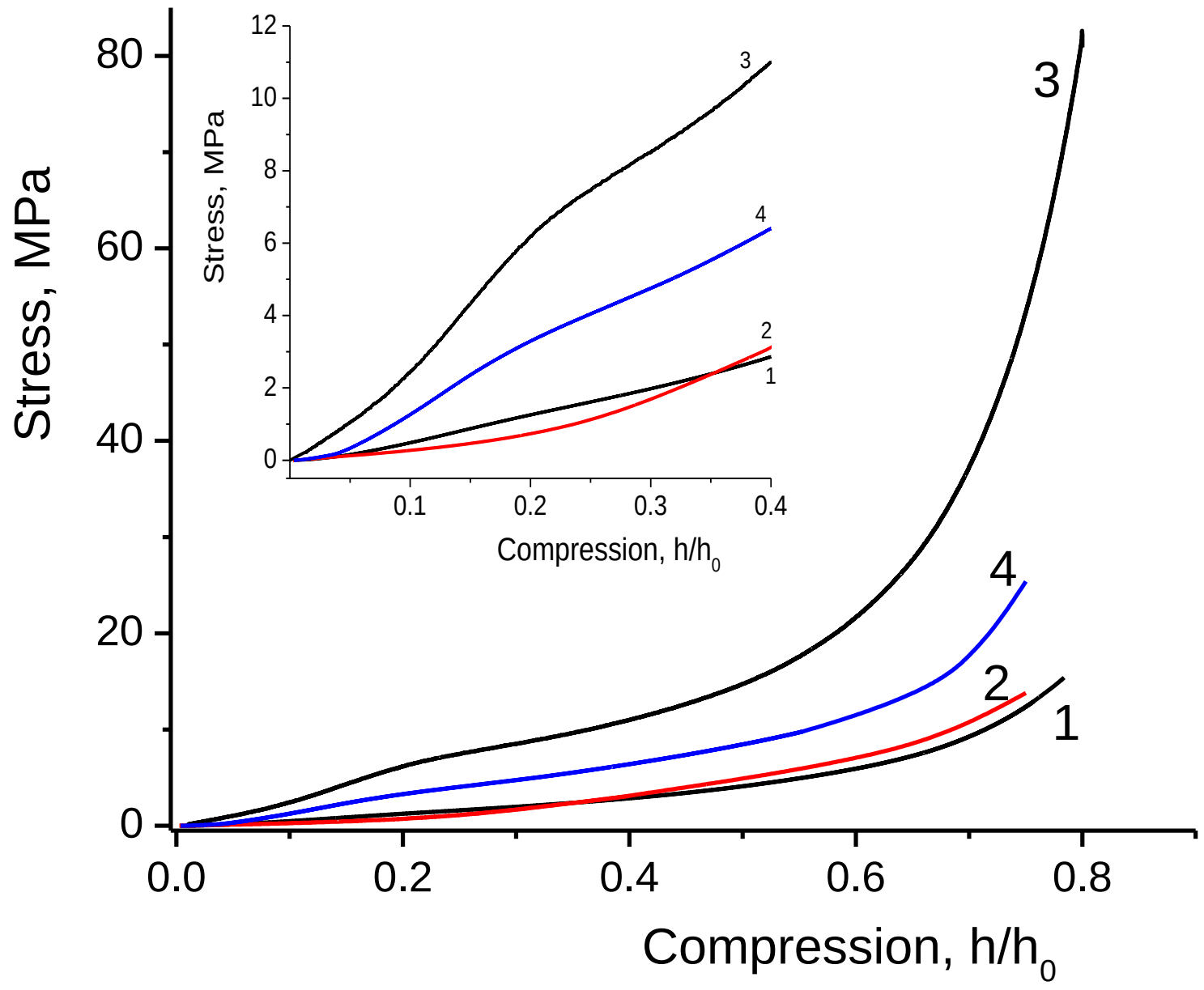

Figure 4. Single compression curves of the samples presented in Table 1. 1 – PC-PAAm containing 7 wt.% cellulose; 2 – PC-PAAm-PAA($Na^+$)_10 containing 7 wt.% cellulose; 3 – PC-PAAm containing 18 wt.% cellulose; 4 – PC-PAAm-PAA($Na^+$)_25 containing 18 wt.% cellulose. The inset shows the initial sections of the same curves (up to the compression equal to 0.4)

To investigate viscoelastic behavior of hydrogels with the stiffness levels typical of natural articular cartilages and to assess their real functional strength, we developed the testing method involving repeated cyclic compression. In these experiments, a cylindrical sample was subjected to a series of compression cycles (10-100 cycles per series, depending on the objectives of the experiment) with a fixed amplitude. After completion of the series of cycles, the experiment was repeated with increased amplitude. Thus, from series to series the amplitude of compression was increased stepwise (from 30 - 40% to 60 - 80%) depending on the level of ultimate breaking strain of a material.

The curves of cyclic compression with the amplitudes of 30, 50, and 70 % for non-ionic sample 1 (Table 1) are displayed in Fig. 5.

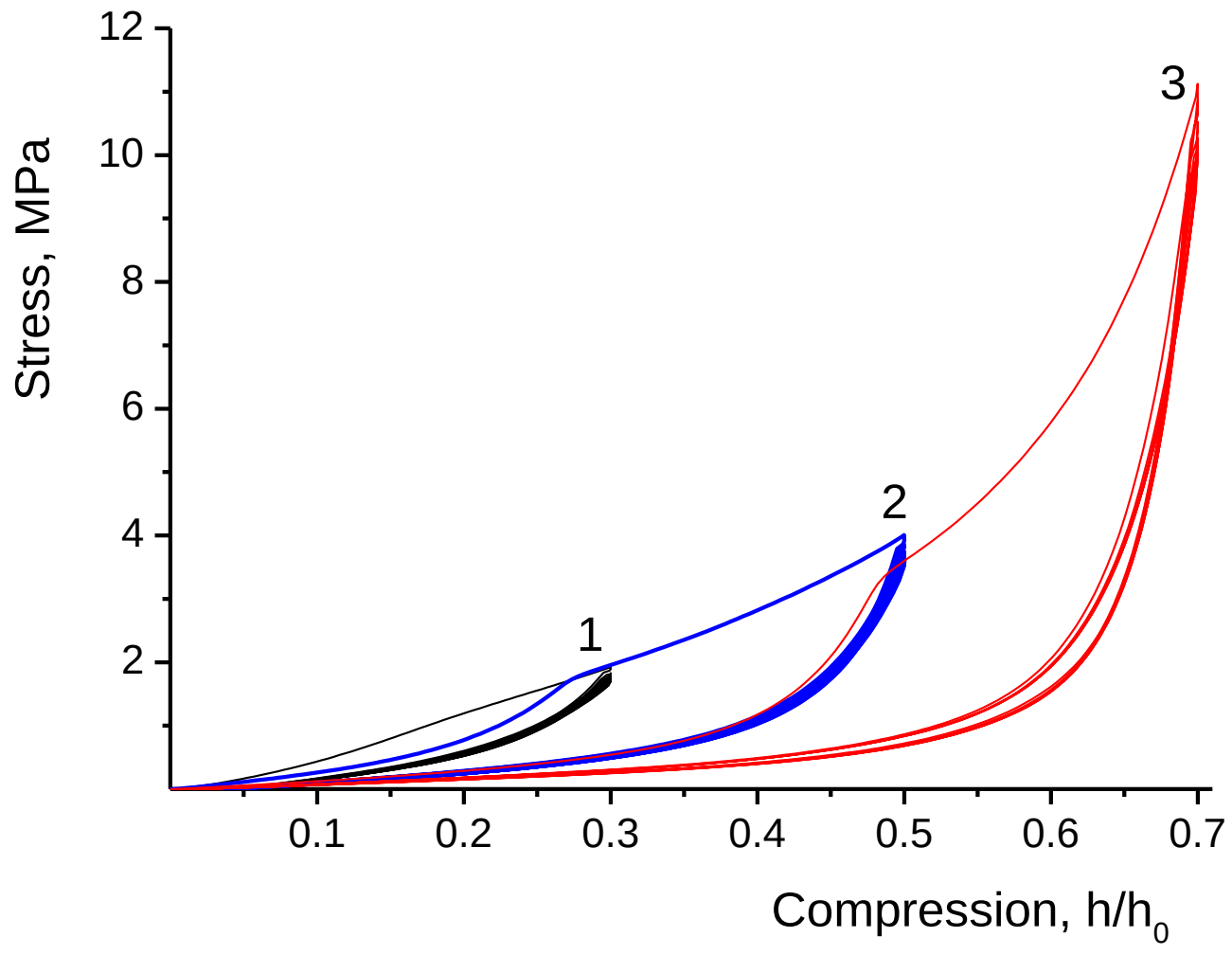

Figure 5. Cyclic compression curves of non-ionic hydrogel 1 (PC-PAAm) obtained at stepwise-increasing compression amplitude (by 20% for each new series of cycles): 50 compression cycles were performed at amplitudes equal to 30 (1), 50 (2), and 70% (3), respectively.

During the first compression cycles of all series (amplitudes of 30, 50, 70 %), a rather wide hysteresis loop is visible. However, already in the second cycle of each series, the hysteresis loop “shrinks” dramatically: the compression curve, which in the first cycle of the series passed noticeably higher than the unloading curve, is located lower and practically overlaps with the unloading curve. At the same time, the compression and unloading branches of the cyclic curves of the second and following cycles coincide with the unloading curve of the first cycle. This means that in the process of compression of the sample in the first cycle (as well as in the case of single compression) serious changes in the structure of the studied materials occur. Apparently, the system of hydrogen bonds that stabilize the IPN structure is being rearranged (which includes breaking of a part of these bonds); possibly, simultaneous rearrangement of the physical network (entanglements of macromolecules) also occurs. However, this hydrogel sample endures multiple cyclic compressions with an amplitude of 70%, and no signs of material failure are observed. The maximum value of compression stresses slightly decreases from cycle to cycle and is close to the value observed in the first compression cycle (11 MPa).

Note that already for the first cycle of the series with amplitudes of 50 and 70 %, the shape of compression curve is determined by the sample history: the compression curve for this cycle coincides with the compression/unloading curves of the second and subsequent cycles of the previous series up to the maximum compression amplitude realized in this previous series (Fig. 5).

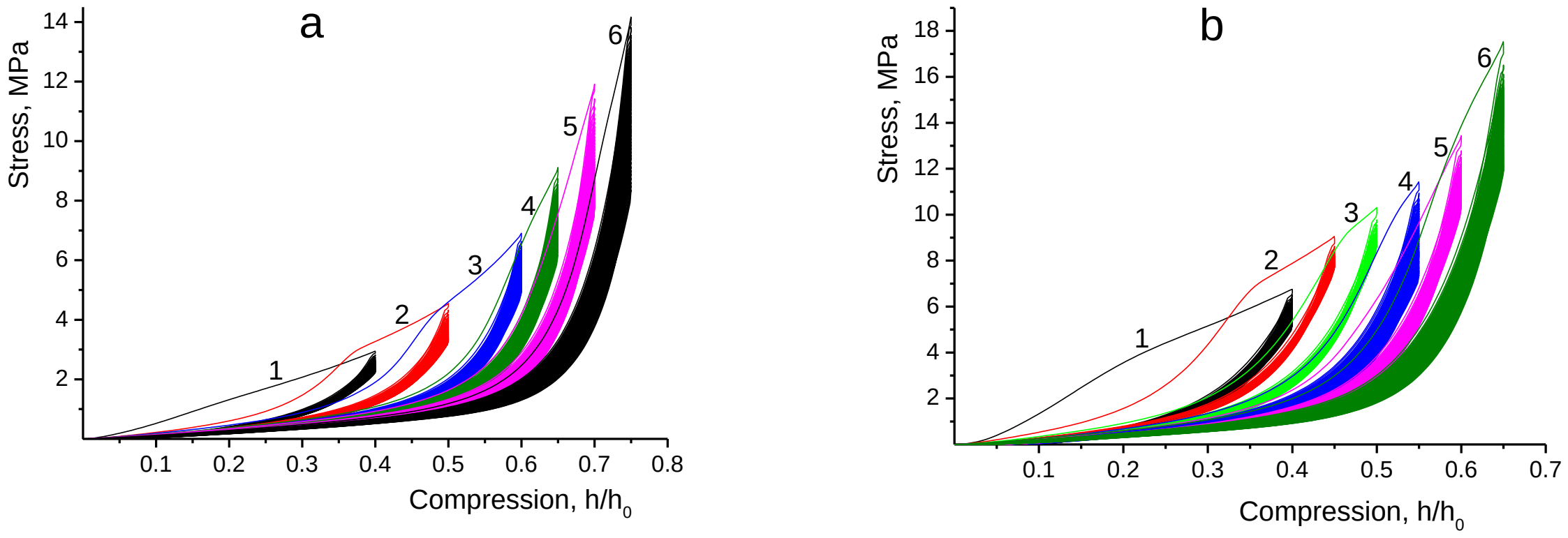

Figure 6. Cyclic compression curves of ionic hydrogels PC-PAAm-PAA($Na^+$)_10 (a) and PC-PAAm-PAA($Na^+$)_25 (b) (samples 2 and 4 in Table 1) obtained at stepwise-increasing compression amplitude (by 5% or 10% for each new series consisting of 50 compression cycles). Compression amplitudes: 1 – 40 %, 2 – 50 %, 3 – 60 %, 4 – 65 %, 5 – 70 %, 6 – 75 % (a);
1 – 40 %, 2 – 45 %, 3 – 50 %, 4 – 55 %, 5 – 60 %, 6 – 65 %, 7 – 70 % (b).

Similar effect is clearly seen in the cyclic compression curves of ionic hydrogels (Fig. 6). Thus, with increasing amplitude, the envelope of all cyclic curves coincides well with the curve of single compression of the material; it has been demonstrated in our earlier work [35] that the maximum values of compression stresses in the cycles of each amplitude become approximately similar to the values observed at single compression of hydrogels. The viscoelastic behavior of the studied hydrogels observed in these experiments has been previously described as a characteristic behavior of elastomers (the Mullins effect [41].

As the compression amplitude increases, a significant reduction in the stiffness level of hydrogel material (a decrease in the slope of the curves) is observed up to high deformation values. In this connection, we proposed to characterize the stiffness of hydrogels by the value of the secant compressive modulus in the deformation range of 10-15 %, which was calculated for one of the compression cycles following the first cycle, for example, the fifth cycle. Table 2 shows the values of secant compressive moduli $E|_{10\text{-}15\%}$ calculated for the fifth cycle compared to the compressive moduli calculated for the first cycle in the same range of strains. Hereinafter, we present the data obtained during the compression cycles with the amplitude of 50%, since we take this value as the maximum amplitude possible during the real functioning of cartilages in joints [17].

Table 2. Comparison of mechanical characteristics of the studied hydrogels measured during single and cyclic compression tests

| Sample No | Sample type | Single compression | | Cyclic compression, the 5th cycle | |
|---|---|---|---|---|---|
| | | $E\|_{10\text{-}15\%}$, MPa | $\sigma\|_{50\%}$, MPa | $E\|_{10\text{-}15\%}$, MPa | $\sigma\|_{50\%}$, MPa |
| 1 | PC-PAAm | 8.14 | 3.99 | 2.28 | 3.92 |
| 2 | PC-PAAm-PAA($Na^+$)_10 | 5.92 | 4.50 | 2.27 | 4.03 |
| 3 | PC-PAAm | 34.9 | 16.6 | 4.92 | 16.78 |
| 4 | PC-PAAm -PAA($Na^+$)_25 | 22.0 | 8.46 | 4.08 | 9.33 |

It can be seen that the values of compressive moduli calculated for the fifth cycle are indeed several times lower than those obtained during single compression of these materials. At the same time, the values of amplitude compressive stresses in the fifth cycles ($\sigma|_{50\%}$) differ only slightly from the values obtained during single compression of the samples.

Taking into account the results of mechanical tests and estimation of equilibrium degrees of swelling, hydrogel samples 1, 2, and 4 (Table 2) were selected for further *in vivo* experiments. In general, the level of mechanical characteristics of the selected hydrogels corresponds to that observed for articular cartilages (when we compare the results obtained by similar methods in the experiments carried out in similar testing regimes).

Since cartilages exhibit viscoelastic properties [42], their mechanical behavior depends on the deformation rate. For example, in [18], it was found that under conditions of unconfined compression, the stress magnitude at 50% deformation for dog knee cartilage increased from approximately 0.7 to 6 MPa as the compression rate increased from 3 to 300 %/min. In our works, the compression rates equal to 1 and 10 mm/min were used, which at 4 mm sample height gives deformation rates of 25 and 250 %/min.

For a more accurate comparison of the discussed values for cartilages and hydrogels, it was advantageous to obtain the dependences of mechanical characteristics on compression rate for the samples studied in this work. The corresponding dependence for sample 1 is shown in Fig. 7.

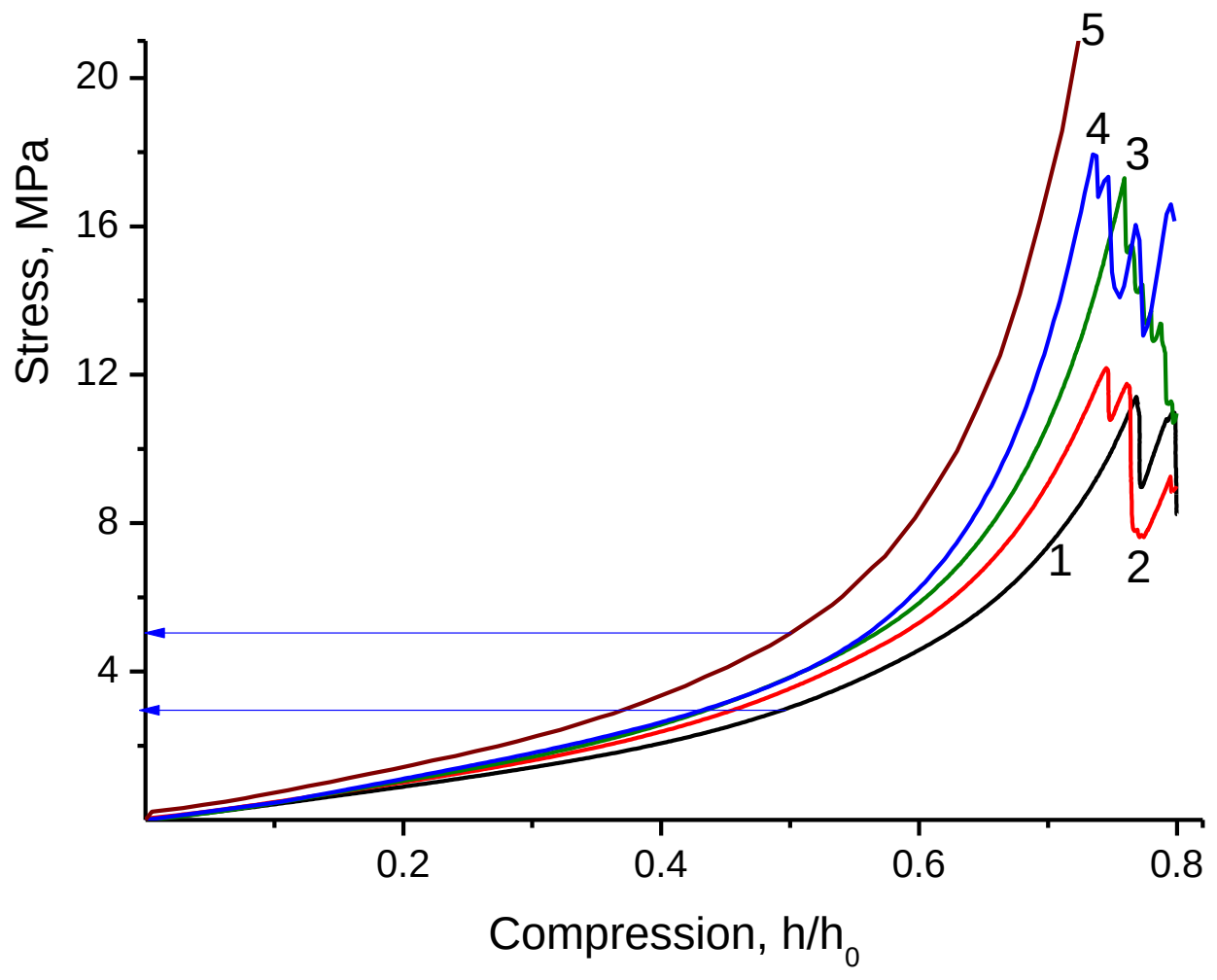

Figure 7. Single compression curves for sample 1 obtained at different deformation rates: 1 – 0.25 %/min; 2 – 2.5 %/min; 3 – 25 %/min; 4 – 250 %/min; 5 – 1250 %/min

At such variation of the deformation rate, very limited changes in the stiffness of the material were registered. Thus, the stress corresponding to compression of the sample by 50% ($\varepsilon = 0.5$) increases from 3.0 to 5.0 MPa with increasing deformation rate over the entire range used in this experiment (Fig. 7). It may be noted that this change in material stiffness turned out to be less significant than that obtained in the experiments involving natural cartilage tissues [18]. This means that for our hydrogel materials, the elastic component contributes much more significantly to the viscoelastic mechanical behavior of the material than in the case of cartilages studied in [18].

However, as was shown in a number of studies of the mechanical behavior of cartilage tissues [18, 43, 44], the variation of their stiffness is quite wide not only for different joints, but also for cartilage regions of different localization within a single joint. Thus, it can be reasonably stated that the deformation behavior of the hydrogel materials studied in this work closely follows the behavior of cartilage tissues.

The experiments carried out in this work showed that the variation of the deformation rate of hydrogel samples within the range of 1-10 mm/min (25-250 %/min) causes only insignificant changes in the characteristics of the material: for example, for sample 1, the compression stress measured at 50% increased from 3.80 to 3.87 MPa. Thus, the variation of rate regimes of hydrogel testing within the specified limits (which in some cases was reasonable for technical reasons) does not lead to any noticeable distortion of the obtained results.

**Mechanical properties of cartilage implants removed from the rabbit's knee 90 - 120 days after operation**

A portion of the hydrogel implants (samples 1, 2, and 4) retrieved after *in vivo* experiments was tested in the cyclic compression regime with increasing amplitude under the same conditions that were used to test the original samples.

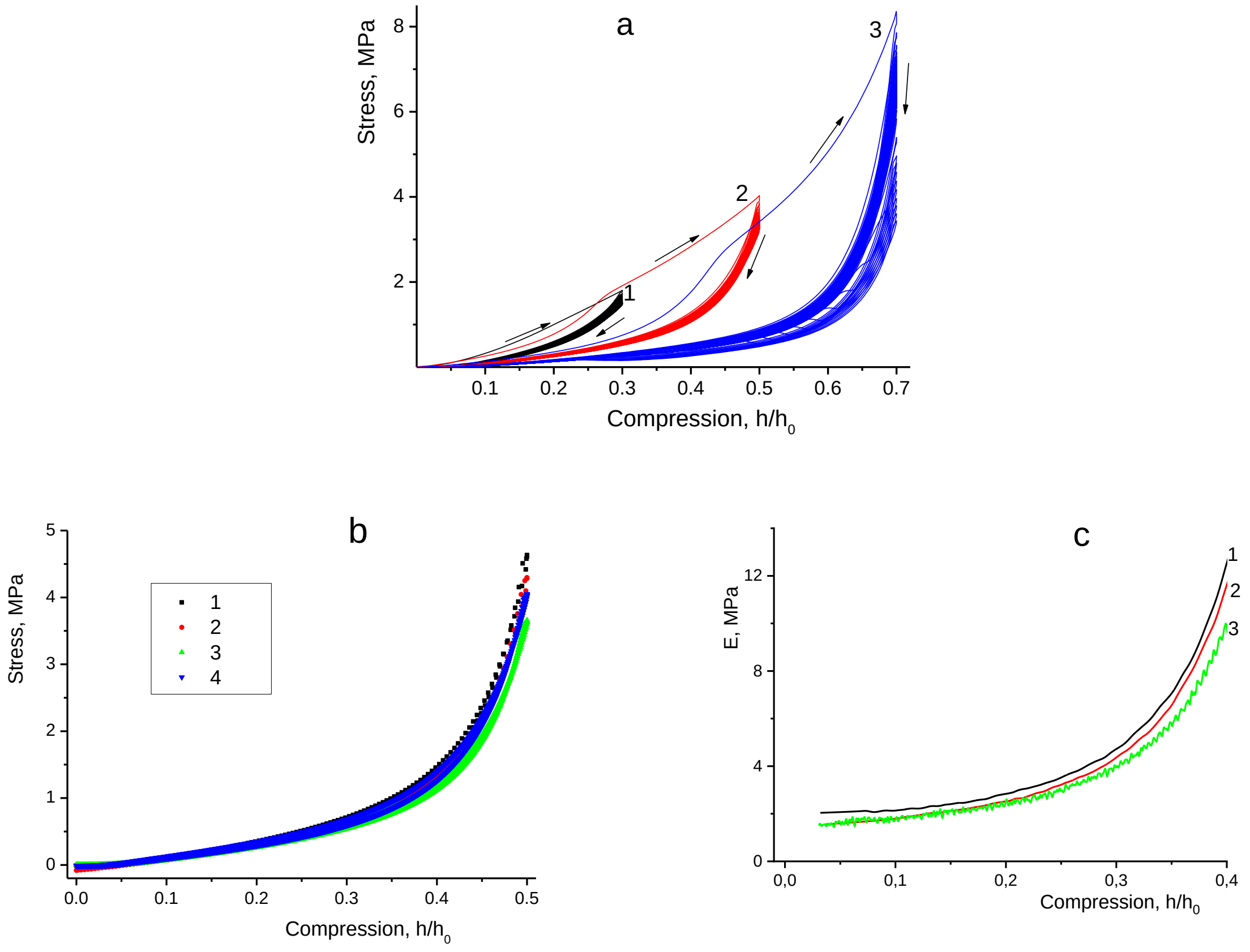

Figure 8. Mechanical characteristics of the implants based on hydrogel 1 measured after functioning in rabbit knee joints as artificial cartilages (loaded and non-loaded areas).

a – cyclic compression curves (100 compression cycles in each series) with amplitudes equal to 30% (series of curves 1), 50% (2) and 70 % (3); the sample was removed from rabbit knee joint (loaded area) 120 days after implantation;

b – curves of fifth cycles of cyclic compression with the amplitude of 50 % of initial sample (1) and the samples removed from joints 90 (2) and 120 (3, 4) days after implantation. 2, 3 – loaded area, 4 – non-loaded area;

c – dependence of the actual compression modulus $d\sigma/d\varepsilon$ on deformation $\varepsilon$ for the samples removed from the loaded areas of joints 90 (2) and 120 days (3) after implantation and for the initial sample (1).

Figure 8a shows the curves of cyclic compression of the implant (non-ionic sample 1) removed from the loaded area of rabbit knee joint 120 days after operation. Compression cycles were applied to the sample with successively increasing amplitude (by 20 %, from 30 to 70 %); 100 compression cycles were performed at each amplitude. It is seen that the viscoelastic behavior of this implant, which was previously observed for the original sample (Fig. 5), is retained. It may be noted that for the cycles with the maximum compression amplitude (70%) there is a certain decrease in the stiffness of the material after its exposure to the joint: the stresses in the compression cycles are somewhat lower than those observed at the same deformations for the original sample (Fig. 5). For the implant subjected to the compression cycles with the amplitude of 70 %, there is a sequential (from cycle to cycle) decrease in the stresses, which was not observed for the initial hydrogel. It may be assumed that in the process of cyclic loading with this high deformation amplitude, the sample began to fracture gradually: during the last cycles, the amplitude value of stress is 2.5 times lower compared to that in the first cycle (Fig. 8a). It is most likely that this behavior is associated with a decrease in the elasticity of the material due to its significant mineralization in the area of contact with the subchondral bone, which leads to the early development of destruction processes in the sample. However, as it was mentioned above, the real compression deformations for cartilages do not exceed 50% even at maximum loads, and at cyclic compression with this amplitude the implant behavior is sufficiently stable (Fig. 8a). The amplitudes of cyclic stress in the material (more than 4 MPa) are close to those for cartilages.

Figure 8b shows that the curves of the fifth compression cycle at an amplitude of 50% for the implants removed 90 and 120 days after placement are very close to the curve for the initial hydrogel (with a slight decrease in stiffness). The most significant decrease in the level of compression stresses (from 4.67 to 3.65 MPa) is observed for the sample extracted from the loaded area of cartilage 120 days after implantation. The curves presenting the dependences of the “actual" compressive modulus $d\sigma/d\varepsilon$ on deformation are also close to each other (Figure 8c). The stiffness level of the material also slightly decreases compared to the initial one for the sample extracted from the loaded area of cartilage 120 days after implantation.

Figure 9 illustrates the results of the analysis of features of mechanical behavior of ionic type implant hydrogels (samples 2 and 4) extracted 90 days after implantation, which are similar to those discussed above for sample 1 (Fig. 8).

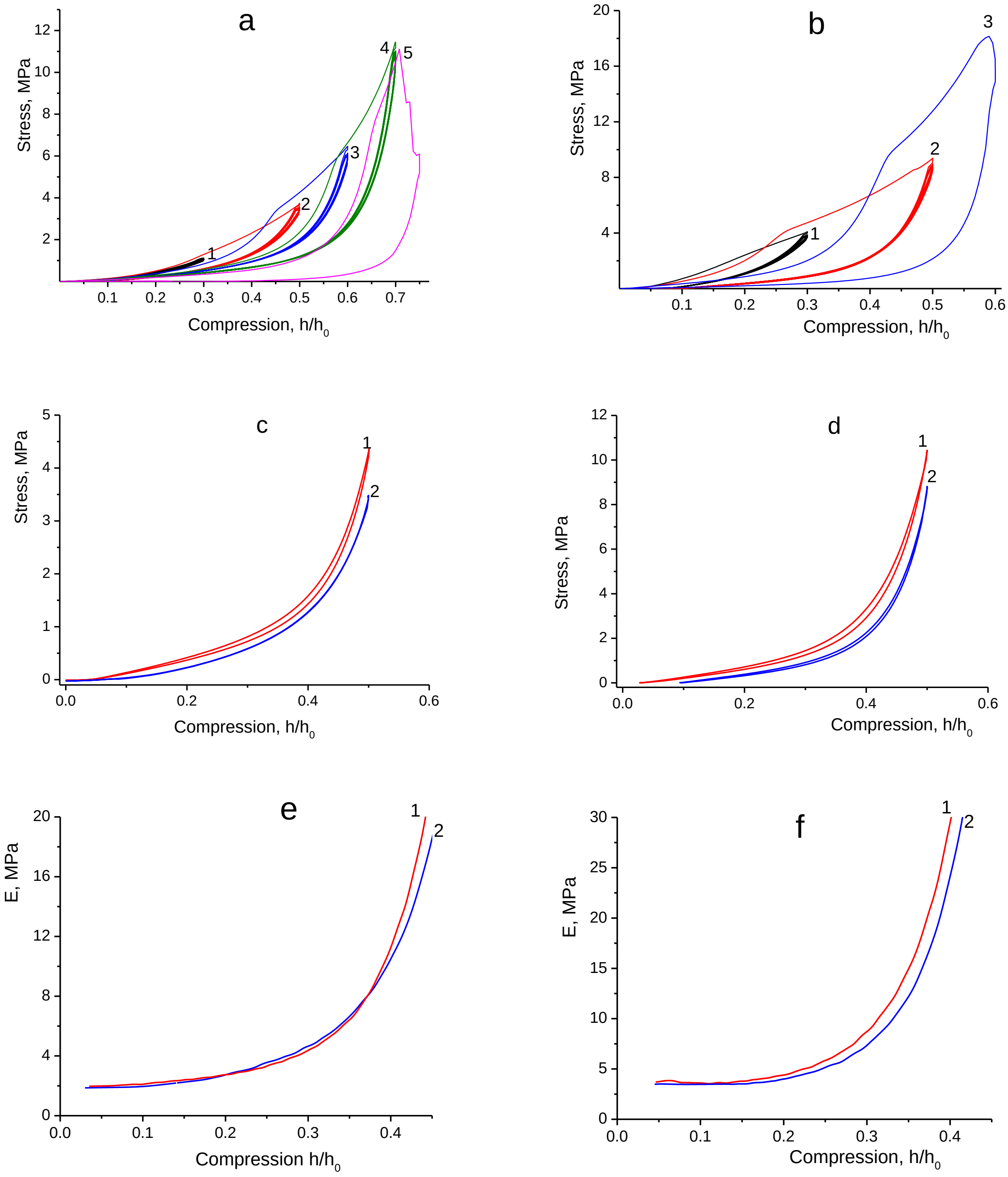

Figure 9. Mechanical characteristics of the implants based on hydrogels 2 (a) and 4 (b) measured after functioning in rabbit knee joints as artificial cartilages (loaded area, 90 days).

a – cyclic compression curves with amplitudes equal to 30% (series of curves 1), 50% (2), 60% (3), 70% (4), and 75% (5); b – cyclic compression curves with amplitudes equal to 30% (series of curves 1), 50% (2), and 60% (3).

c, d – curves of fifth cycles of cyclic compression with an amplitude of 50 % of samples 2 and 4 removed from joints, respectively (curves 2) compared to those for the corresponding initial samples (curves 1);

e, f – dependences of the actual compression modulus $d\sigma/d\varepsilon$ on deformation for samples 2 and 4 removed from joints, respectively (curves 2) compared to those for the corresponding initial samples (curves 1).

The implant sample prepared of ionic hydrogel 2 withstood the compression cycles at the amplitude up to 70 % inclusive, but upon increase in the amplitude up to 75 % it was destroyed during the first compression cycle (Fig. 9a, curve 5). The stiffer sample of ionic hydrogel 4 withstood the effects of compression cycles at the amplitude of 50 %, but collapsed during compression in the first cycle when the amplitude was increased up to 60 %. On the contrary, the initial hydrogel samples 2 and 4 showed higher elasticity and tolerated cyclic compressions up to amplitudes of 75 and 65 %, respectively (Fig. 6).

Thus, we observe a certain loss of elasticity, i.e. the aging process occurring during functioning in the joint for both non-ionic and ionic types of implants. These processes are intensively manifested at increasing amplitude of compression, namely at amplitudes above 50%. Most likely, this behavior is related to the decrease in the elasticity of a material due to its mineralization in the area of contact with the subchondral bone. At the same time, as it was mentioned above, the compressive deformations to which cartilage in joints is actually subjected, even under maximum loads, do not exceed 50%. Therefore, a drop in compressive strain at the amplitude of more than 50% is not critical.

A certain decrease in the level of compression stresses of the extracted implants in comparison with the original samples is observed for the fifth compression cycles at the amplitude of 50 % (Fig. 9c and 9d for samples 2 and 4, respectively). However, such changes are also tolerable, since the implant retained a sufficiently high level of compression loads.

The dependences of the "actual" compressive modulus $d\sigma/d\varepsilon$ on the deformation value for the extracted ionic implants are very close to the corresponding curves for the initial samples (Fig. 9e and 9f). A similar behavior of non-ionic hydrogels is demonstrated in Fig. 8c.

Thus, the data discussed above showed that during the long-term functioning of hydrogels as artificial cartilages in the joints of animals (up to 90 - 120 days) there was no critical deterioration in the level of mechanical characteristics of these materials.

## SEM and EDX studies of chemical composition and morphology of hydrogel implants

### Non-ionic hydrogels

It has been shown [38] that the area of non-ionic cellulose/PAAm hydrogel implants located in the subchondral bone undergoes mineralization with the formation of calcium phosphate clusters in the gel volume. Chemical composition of these clusters is close to that of hydroxyapatite. Thus, it was desirable to obtain more detailed data on the morphology and structure of both non-ionic and ionic implants by SEM and to estimate the calcium phosphate content in the implants by EDX.

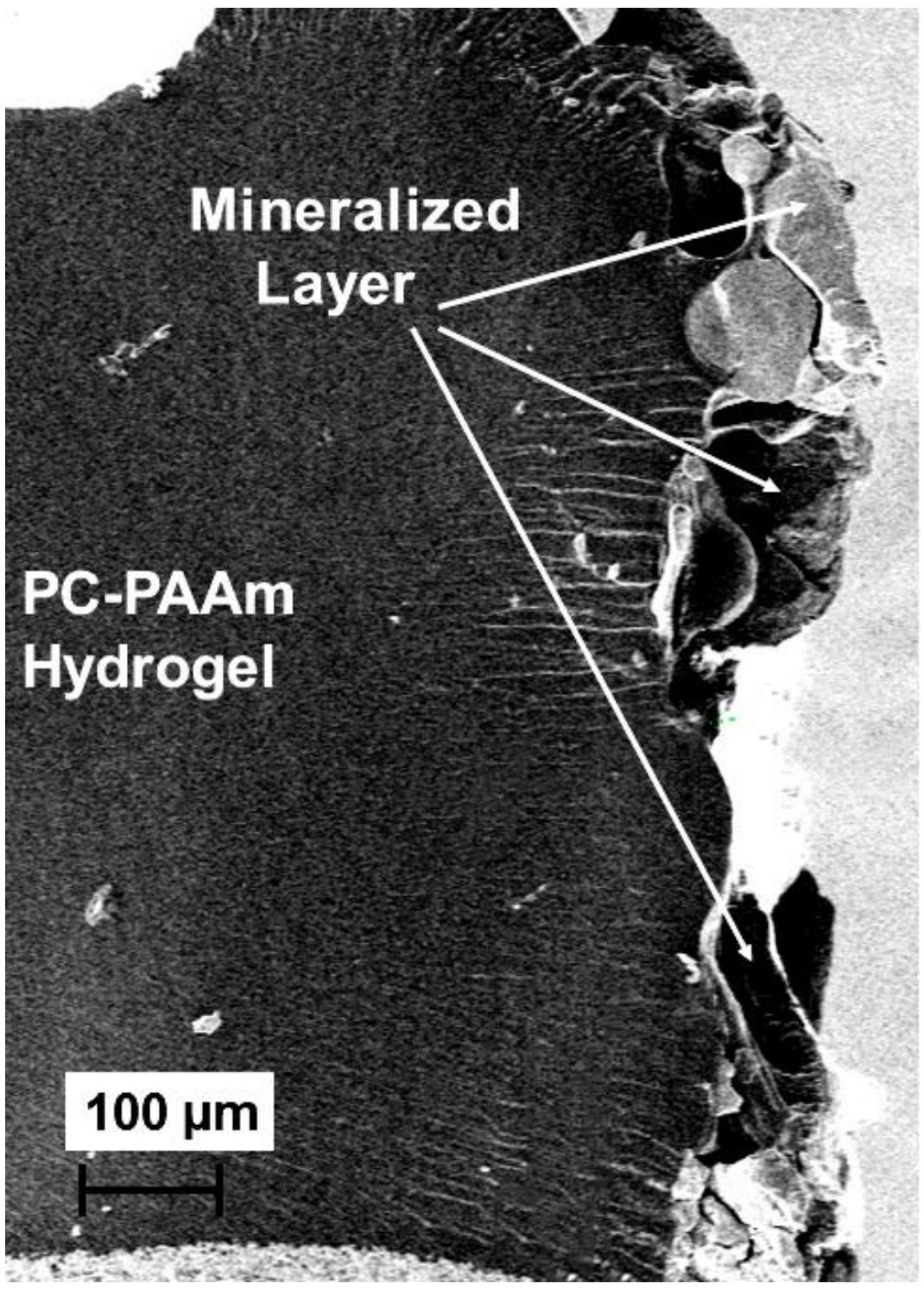

Figure 10. SEM microphotograph of the fractured surface of the subchondral area of a hydrogel implant (the initial sample 1) extracted from the rabbit body after 120 days of the *in vivo*. Magnification ×200.

The fractured surface of implant 1 that was removed from the rabbit organism after 120 days of the *in vivo* experiment is displayed in Fig. 10. The main part of the implant has a predominantly homogeneous structure, but in its right part, a non-uniform layer with a thickness of about 100 μm can be seen. Separate parts of this layer were investigated by SEM and EDX. This layer is located on the lateral surface of the cylindrical implant sample, which contacted with the subchondral bone.

The SEM image of one region of this layer taken at higher magnification (×500) is shown in Fig. 11. Both local EDX scanning points and two rather large scanning areas marked by rectangles (spectra No. 3 and No. 6) are shown. Scanning area No 6 records high contents of both calcium and phosphorus: 18.3 and 9.05 wt.%, respectively (Table 3), giving about 27 wt.% calcium phosphate in the composition of the material at the studied area. At the same time, about 5.3 wt.% of nitrogen is found; that is, calcium phosphate in this zone is embedded in the polymer network of the hydrogel. The initial hydrogel contained about 20.1 wt.% nitrogen, which is characteristic of formulations containing predominantly PAAm. Scanning region No 6 also recorded a reduced carbon content (25.1 wt.%), but the content of oxygen (41.9 wt.%), which is incorporated into the calcium phosphates, is much higher than that in the initial hydrogel (24.3 wt.%).

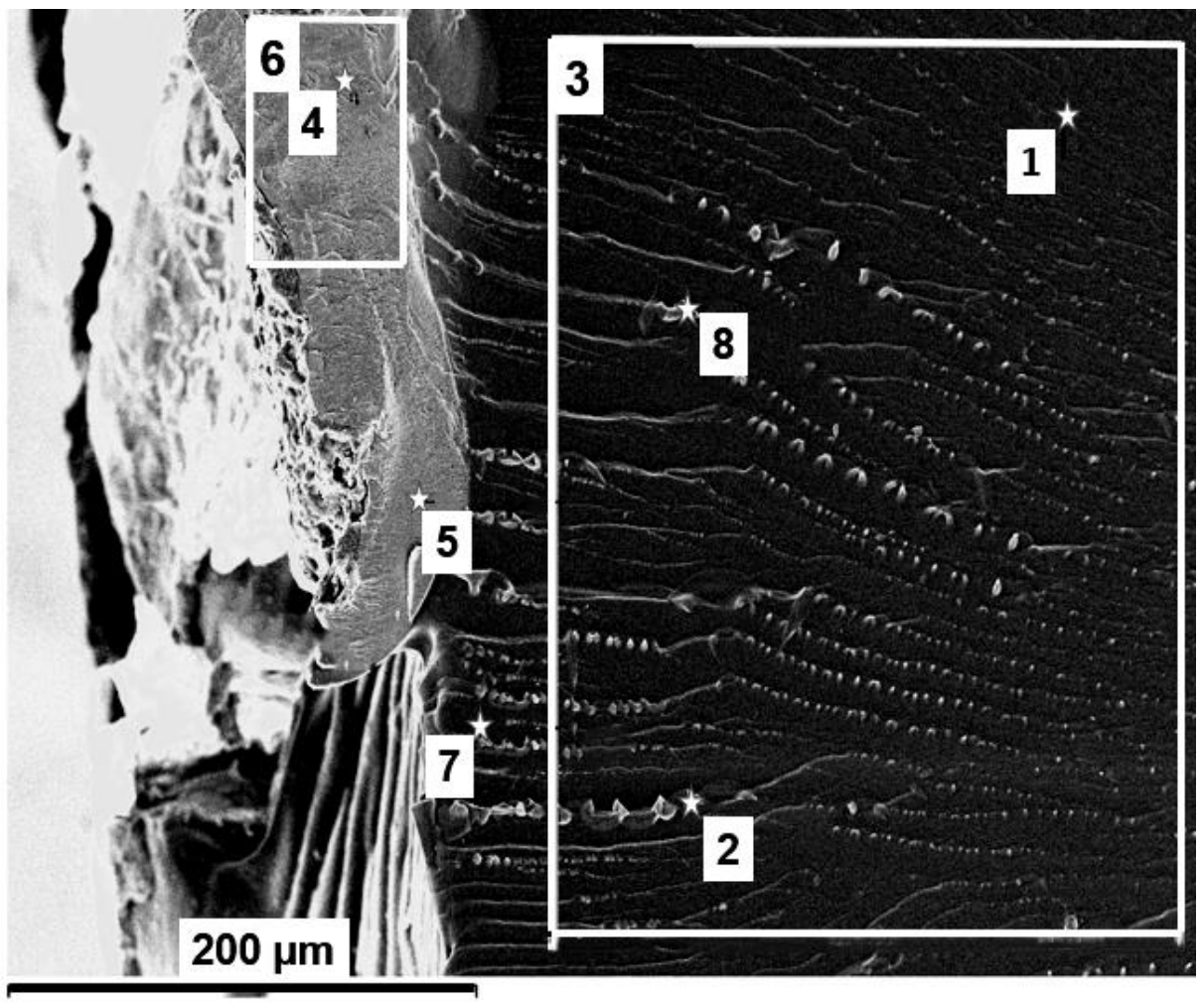

Figure 11. SEM microphotograph of the fractured surface of the subchondral area of a hydrogel implant (the initial sample 1) extracted from the rabbit body after 120 days of the *in vivo* experiment. Loaded area of cartilage. Magnification ×500.

The calcium content inside the large rectangular area (spectrum 3, Fig. 11) is very low (0.15 wt.%), and the phosphorus content is equal to 0. Meanwhile, the contents of other elements (C, N, O) are close to those in the initial implant. Thus, it may be stated that the mineralization process occurred inside the local layer of the hydrogel implant, beyond which signs of mineralization are completely absent. In individual scanning points 1, 2 and 8, which are located outside the boundary of the mineralized area of the implant, the contents of elements are close to those observed for scanning area 3.

Table 3. Chemical composition of the initial implant and its “subchondral” area after the *in vivo* experiment at the points and scanning areas indicated in Fig. 11. Here and in the following tables, the concentrations of elements are given in wt.%.

| Spectrum No | Concentrations of elements, wt.%. | | | | | |
|---|---|---|---|---|---|---|
| | C | N | O | Na | P | Ca |
| Initial implant | | | | | | |
| | 55.59 | 20.13 | 24.28 | 0 | 0 | 0 |
| Implant after *in vivo* experiment | | | | | | |
| 1 | 62.32 | 16.64 | 20.79 | 0 | 0.00 | 0.25 |
| 2 | 57.80 | 20.38 | 21.57 | 0 | 0.00 | 0.25 |
| 3 | 51.94 | 20.39 | 27.52 | 0 | 0.00 | 0.15 |
| 4 | 26.04 | 5.98 | 30.78 | 0.27 | 12.35 | 24.58 |
| 5 | 23.88 | 1.31 | 37.31 | 0 | 12.88 | 24.62 |
| 6 | 25.10 | 5.28 | 41.96 | 0.29 | 9.05 | 18.32 |
| 7 | 65.33 | 14.61 | 19.95 | 0 | 0.00 | 0.11 |
| 8 | 60.49 | 17.57 | 21.77 | 0 | 0.00 | 0.17 |

A more thorough analysis of the changes in the contents of elements in the implant was carried out in the vicinity of the highlighted area No. 6 (Fig. 11), along the line running through the mineralized and non-mineralized areas of the fractured surface of implant 1 with the scanning points 1 - 15, located with a step of about 10 μm in the direction going into the depth of the implant (Fig. 12, Table 4).

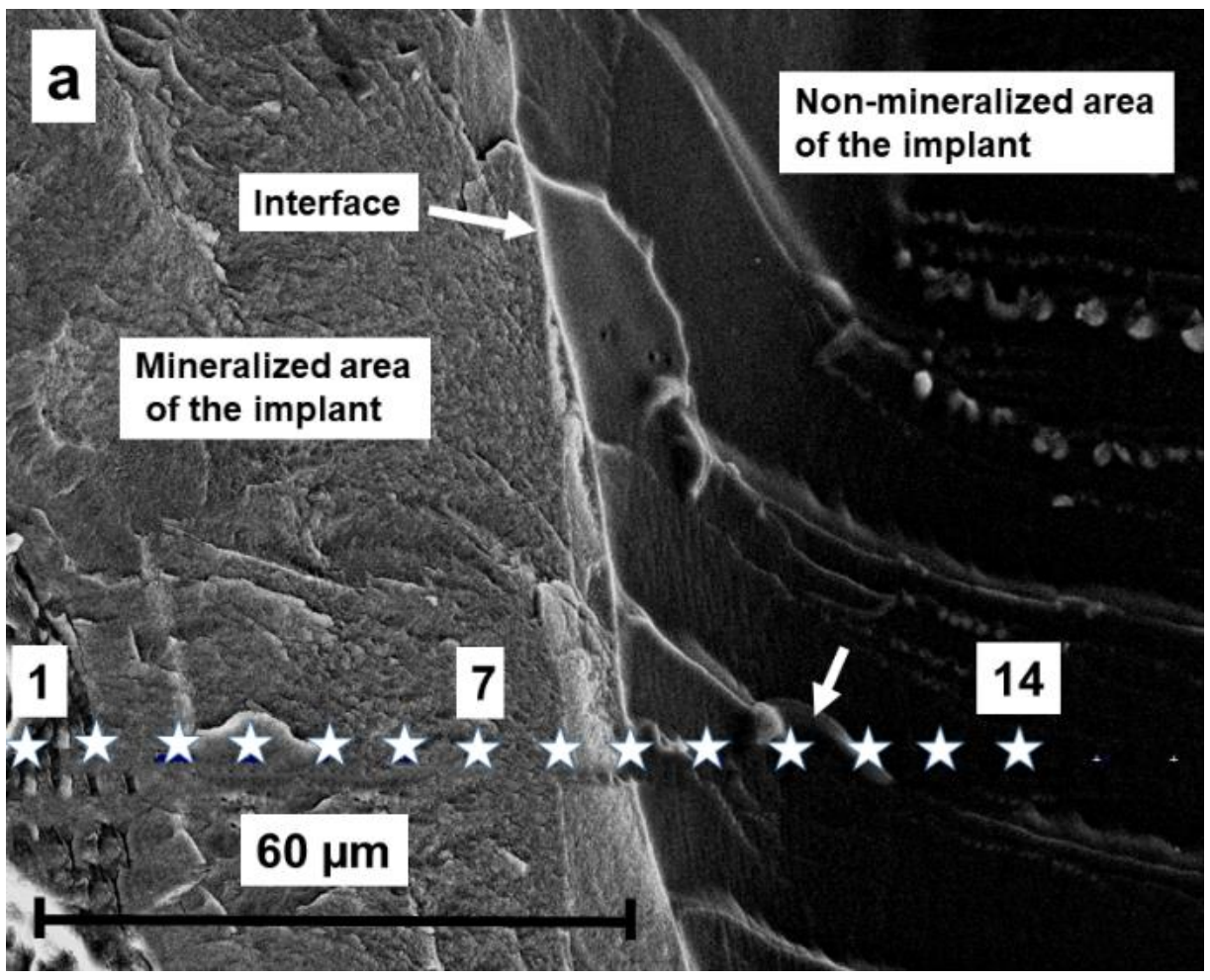

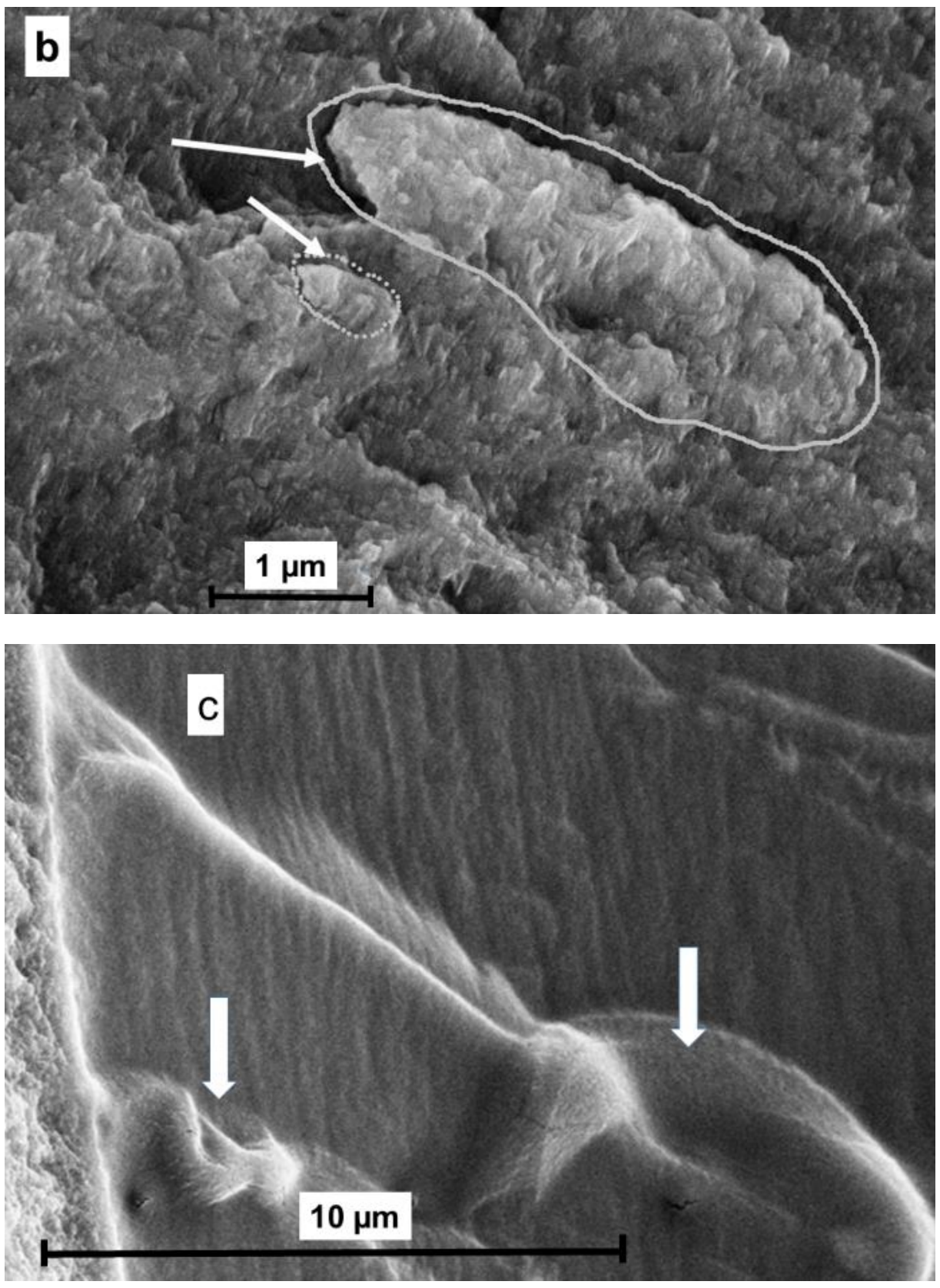

Figure 12. SEM microphotographs of the fractured surface of the subchondral areas of hydrogel implants (the initial sample 1) extracted from the rabbit body after 120 days of the *in vivo* experiment (a), the mineralized area of the implant at higher magnification (b), fragment of Fig. 12a marked with an arrow (c). Loaded area of cartilage. Magnifications ×5000 (a), ×50000 (b) and ×20000 (c).

There is a clear boundary between the mineralized area of the implant, which contains significant amounts of calcium phosphate (up to 40 wt.%, see Table 4), and the area almost completely free of this substance. The Ca/P ratio in the first area (1.92 - 2.09) is close to the characteristic ratio for synthetic hydroxyapatite $Ca_5(PO_4)_3OH$ – 2.16. The presence of carbon and nitrogen in scanning points 2-8 indicates that the mineral phase is embedded into the organic polymeric phase, i.e., into the polymeric network of the hydrogel. Two implant regions differing in morphology with the defined interface between them (Fig. 12a) fit quite well to each other.

Table 4. Chemical composition of the "subchondral" area of the implants in the scanning points indicated in Fig. 12a.

| Spectrum No | Concentrations of elements, wt.%. | | | | | | |
|---|---|---|---|---|---|---|---|
| | C | N | O | Na | P | Ca | Ca/P |
| 1 | 20.03 | 2.26 | 37.90 | 0.32 | 13.05 | 26.44 | 2.03 |
| 2 | 34.76 | 10.82 | 30.89 | 0.14 | 8.01 | 15.38 | 1.92 |
| 3 | 19.73 | 2.27 | 36.19 | 0.32 | 14.05 | 27.44 | 1.95 |
| 4 | 16.54 | 1.99 | 43.40 | 0.28 | 12.25 | 25.54 | 2.09 |
| 5 | 22.61 | 1.66 | 35.65 | 0.39 | 13.23 | 26.46 | 2.00 |
| 6 | 18.59 | 1.83 | 33.09 | 0.35 | 15.11 | 31.03 | 2.05 |
| 7 | 22.75 | 1.90 | 38.64 | 0.35 | 12.21 | 24.15 | 1.98 |
| 8 | 62.41 | 17.68 | 18.99 | 0.00 | 0.22 | 0.70 | - |
| 9 | 61.08 | 18.53 | 19.86 | 0.00 | 0.00 | 0.53 | - |
| 10 | 61.84 | 16.83 | 20.93 | 0.00 | 0.00 | 0.40 | - |
| 11 | 60.77 | 17.85 | 21.09 | 0.00 | 0.00 | 0.29 | - |
| 12 | 62.56 | 17.29 | 19.65 | 0.00 | 0.00 | 0.50 | - |
| 13 | 64.03 | 15.51 | 20.07 | 0.08 | 0.00 | 0.31 | - |
| 14 | 62.47 | 17.01 | 20.15 | 0.00 | 0.00 | 0.37 | - |

The integration between the mentioned areas of the implants was not disturbed even during cyclic tests up to high degrees of compression. It should be also noted that chemical composition and structure of the mineralized areas of the implants are close to those of bone tissues, and the formation of this mineralized material probably reflects the process of the integration of the implant with living tissues. This assumption is indirectly confirmed by the presence of certain formations with the size up to 10 μm in both implant zones, which can be either traces of cells or their waste products. One of these formations is pointed by arrow in Fig. 12a, and in Fig. 12c a fragment of this drawing is shown in close-up. It is likely that osteoblasts penetrate into the surface layers of implant hydrogels and start to actively form hydroxyapatite clusters in these layers.

Fig. 12b displays the mineralized area of the implant at a magnification of ×50 000. Calcium phosphate crystals with the size of 100 nm and higher can be seen; these crystals form much larger clusters with the size reaching several microns (indicated in the image by arrows).

Along the implant-subchondral bone interface, crystalline formations with a certain ordered structure are observed in some places (Fig. 13). Ring-shaped layers are visible around the scanning point labeled with the number 1. In all six scanning points, phosphorus and calcium are present in concentrations up to 18 and 44 wt.%, respectively, but the Ca/P ratio is much higher than 2 (2.34 - 2.52); in two points this value even reaches 3.5 and 9.0 (points 4 and 6, respectively).

Thus, these crystalline regions apparently contain calcium phosphates differing in composition from hydroxyapatite. Judging from the presence of nitrogen in all spectra in the amounts ranging from 2.2 to 8 wt.%, we can conclude (similarly to the results discussed earlier) that in Fig. 13 we observe not purely mineral formations, but most likely polymer-inorganic objects, i.e. calcium phosphate is formed inside the polymer network of the hydrogel implant.

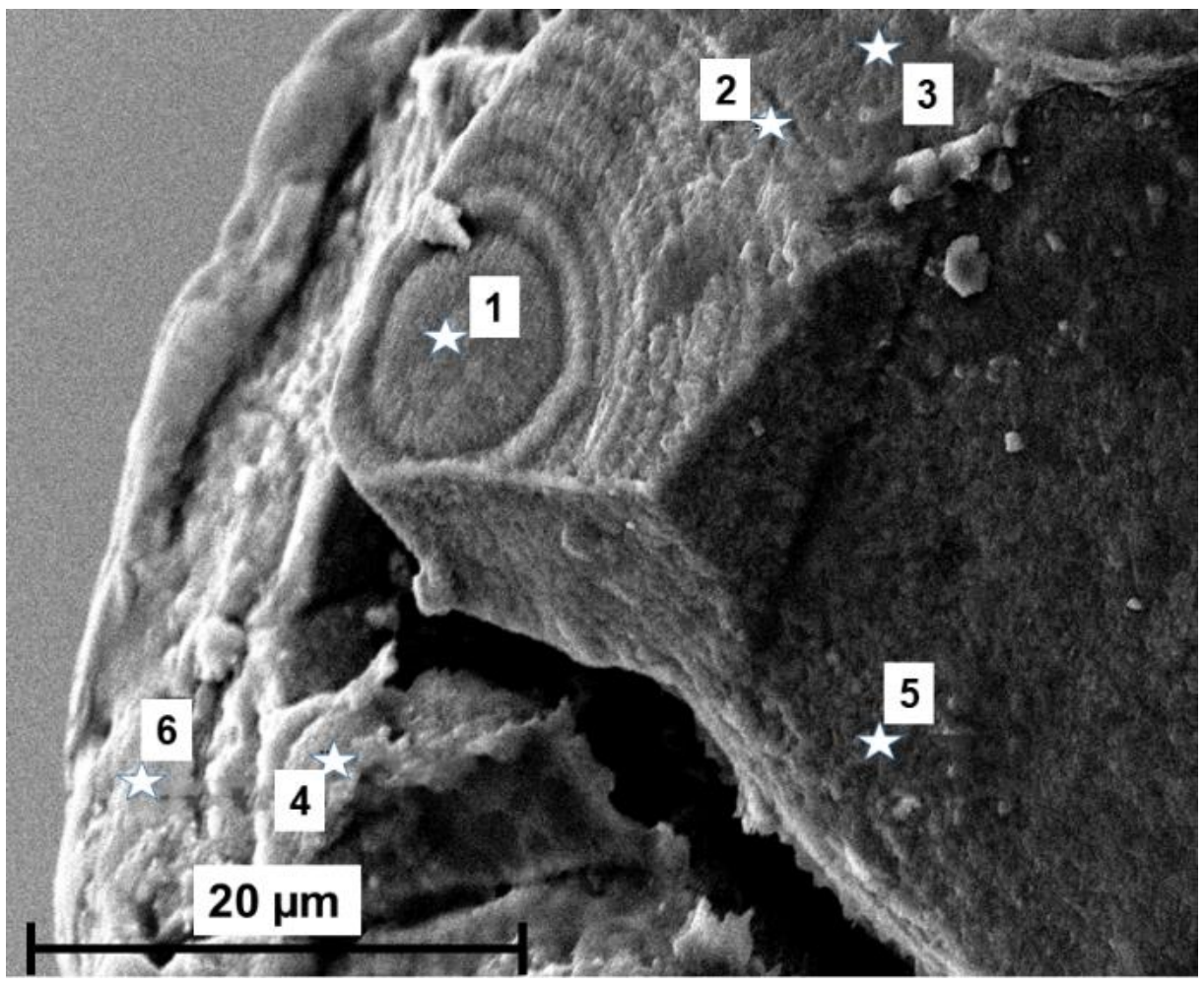

Figure 13. SEM microphotograph of one of the areas of mineralized layer of the implant extracted from the rabbit body after 120 days of the *in vivo* experiment. Loaded area of cartilage. Magnification ×5000.

Table 5. Chemical composition of one of crystalline areas of the implant in the scanning areas indicated in Fig. 10.

| Spectrum No | Concentrations of elements, wt.%. | | | | | | |
|---|---|---|---|---|---|---|---|
| | C | N | O | Na | P | Ca | Ca/P |
| 1 | 20.32 | 3.65 | 14.86 | 0.36 | 17.29 | 43.52 | 2.52 |
| 2 | 22.99 | 2.46 | 26.14 | 0.30 | 14.15 | 33.96 | 2.34 |
| 3 | 16.40 | 2.24 | 18.86 | 0.24 | 17.88 | 44.38 | 2.48 |
| 4 | 23.73 | 3.82 | 17.21 | 0.16 | 12.32 | 42.76 | 3.47 |
| 5 | 22.45 | 7.95 | 32.01 | 0.21 | 10.64 | 26.74 | 2.51 |
| 6 | 32.94 | 3.53 | 22.04 | 0.18 | 4.13 | 37.18 | 9.00 |

Thus, during 120 days of exposure of the non-ionic implant to the subchondral zone of a cartilage defect, the thickness of its mineralized area reaches about 100 µm (Figs. 10 and 11). The characteristics of the hydrogel exposed to the loaded area of rabbit knee joint cartilage obtained by SEM and EDX do not differ principally from those reported earlier for a similar implant extracted from the non-loaded area of the cartilage after 45 days of *in vivo* experiment [38]. In both cases, the average calcium phosphate content in the mineralized zone reaches 35 - 40 wt.%, and the Ca/P ratio is close to 2, which is typical of hydroxyapatite. Moreover, the SEM studies revealed no fundamental differences in the morphology of implants extracted from the loaded (this work) and non-loaded regions of cartilage [38]. In view of this observation, further studies were focused on the implants extracted from the loaded areas of cartilage.

**Ionic hydrogels**

The fractured surfaces of ionic sample 2 (Fig. 14) were obtained and tested in the same manner as those of non-ionic implant 1; elemental composition of this sample was determined by EDX in scanning points 1 – 16 (Table 6).

In this ionic implant, significant amounts of phosphorus and calcium are detected only in three scanning points (8, 12 and 16); they are not located near the lateral surface of the implant, which on the fractured surface runs along the right part of Fig. 14. In points 1 – 6 (near the lateral surface of the implant), the contents of these elements are insignificant: 0.8 – 2.46 wt.% of calcium and only 0.12 – 0.37 wt.% of phosphorus.

In other scanning points, the contents of calcium and phosphorus are also low in comparison with those observed for non-ionic implant 1. It can be stated that in this ionic implant, the mineralization process, as it was supposed, develops much slower than in the non-ionic implant. This conclusion is confirmed by the data for ionic implant 4 presented below.

To study implant 4 by SEM and EDX, hydrogel samples were excised from the joints together with the surrounding bone and cartilage tissue fragments. In Fig. 15, the implant boundaries are indicated by arrows, and the bone tissue surrounding the implant can be seen in the left and lower parts of the image.

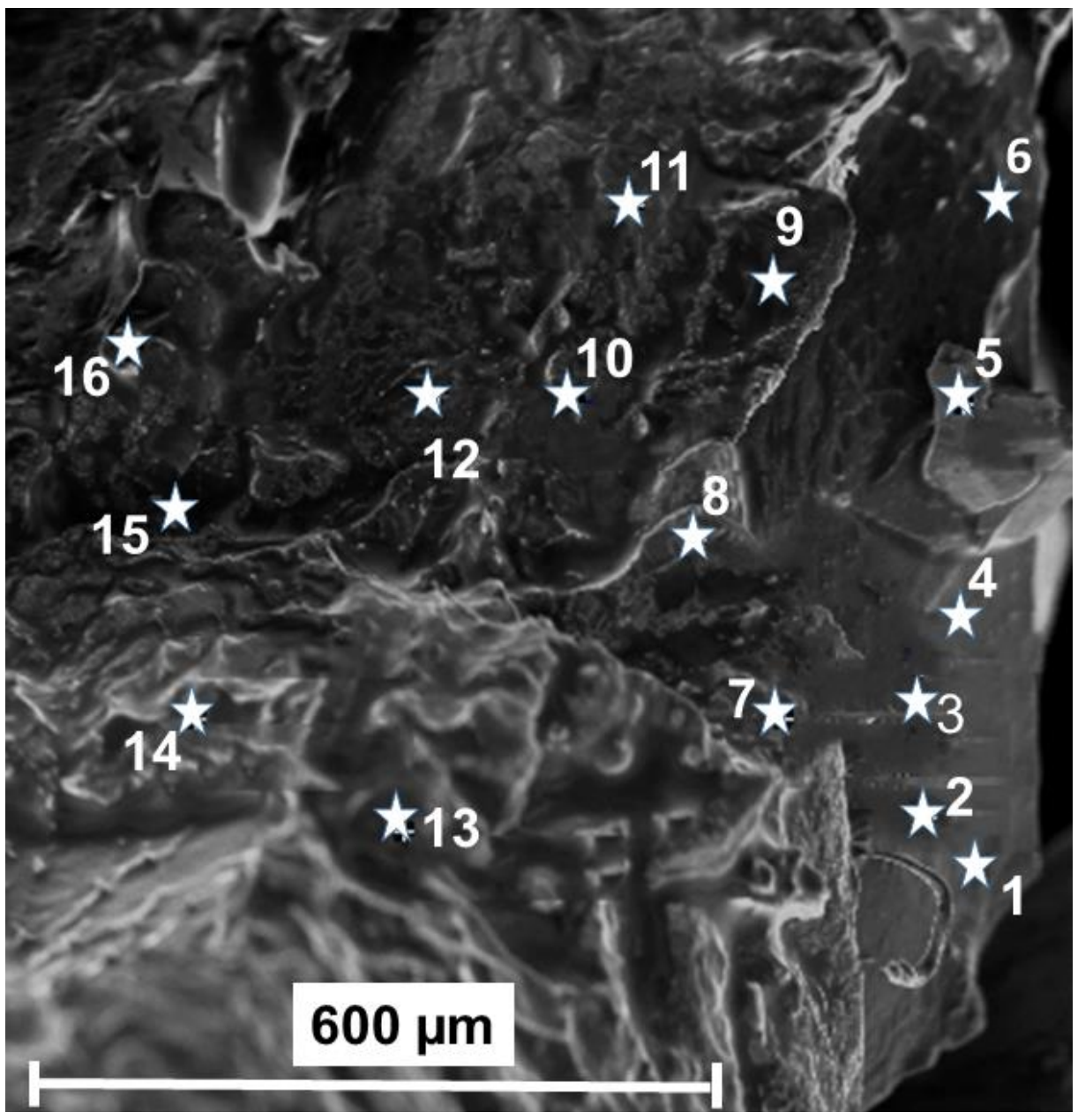

Figure 14. SEM microphotograph of the fractured surface of the subchondral area of implant 2 (near the lateral surface of the sample) extracted from the rabbit body after 120 days of the *in vivo* experiment. Loaded area of cartilage. Magnification ×200.

Table 6. Chemical composition of implant 2 at the scanning points indicated in Fig. 14.

| Spectrum No | Concentrations of elements, wt.%. | | | | | | | |
|---|---|---|---|---|---|---|---|---|
| | C | N | O | Na | Mg | P | S | Ca |
| | Initial implant | | | | | | | |
| | 64.08 | 14.64 | 21.28 | 0.00 | 0.00 | 0.00 | 0.00 | 0.00 |
| | Implant after *in vivo* experiment | | | | | | | |
| 1 | 64.04 | 14.69 | 20.98 | 0.05 | 0.00 | 0.21 | 0.03 | 0.80 |
| 2 | 64.34 | 14.54 | 20.36 | 0.00 | 0.05 | 0.12 | 0.02 | 0.57 |
| 3 | 63.64 | 15.08 | 19.68 | 0.03 | 0.00 | 0.27 | 0.00 | 1.30 |
| 4 | 65.53 | 25.67 | 16.01 | 0.01 | 0.00 | 0.22 | 0.10 | 1.46 |
| 5 | 61.72 | 17.26 | 18.09 | 0.09 | 0.00 | 0.37 | 0.01 | 2.46 |
| 6 | 63.14 | 14.62 | 20.13 | 0.00 | 0.04 | 0.35 | 0.07 | 1.65 |
| 7 | 60.65 | 15.51 | 20.54 | 0.00 | 0.12 | 0.50 | 0.12 | 2.56 |
| 8 | 52.54 | 10.81 | 28.06 | 0.21 | 0.07 | 2.98 | 0.05 | 5.28 |
| 9 | 55.26 | 20.44 | 22.09 | 0.15 | 0.05 | 0.64 | 0.32 | 1.05 |
| 10 | 65.32 | 13.43 | 16.69 | 0.06 | 0.05 | 1.64 | 0.13 | 2.68 |
| 11 | 48.40 | 14.53 | 30.98 | 0.06 | 0.10 | 0.13 | 0.11 | 5.69 |
| 12 | 57.60 | 11.67 | 15.75 | 0.08 | 0.11 | 4.45 | 0.10 | 10.24 |
| 13 | 56.73 | 22.64 | 19.47 | 0.13 | 0.05 | 0.00 | 0.66 | 0.35 |
| 14 | 61.69 | 16.86 | 20.22 | 0.10 | 0.02 | 0.00 | 0.02 | 1.08 |
| 15 | 65.29 | 14.10 | 20.32 | 0.11 | 0.03 | 0.00 | 0.02 | 0.13 |
| 16 | 40.45 | 14.77 | 35.11 | 0.18 | 0.14 | 3.65 | 0.11 | 5.59 |

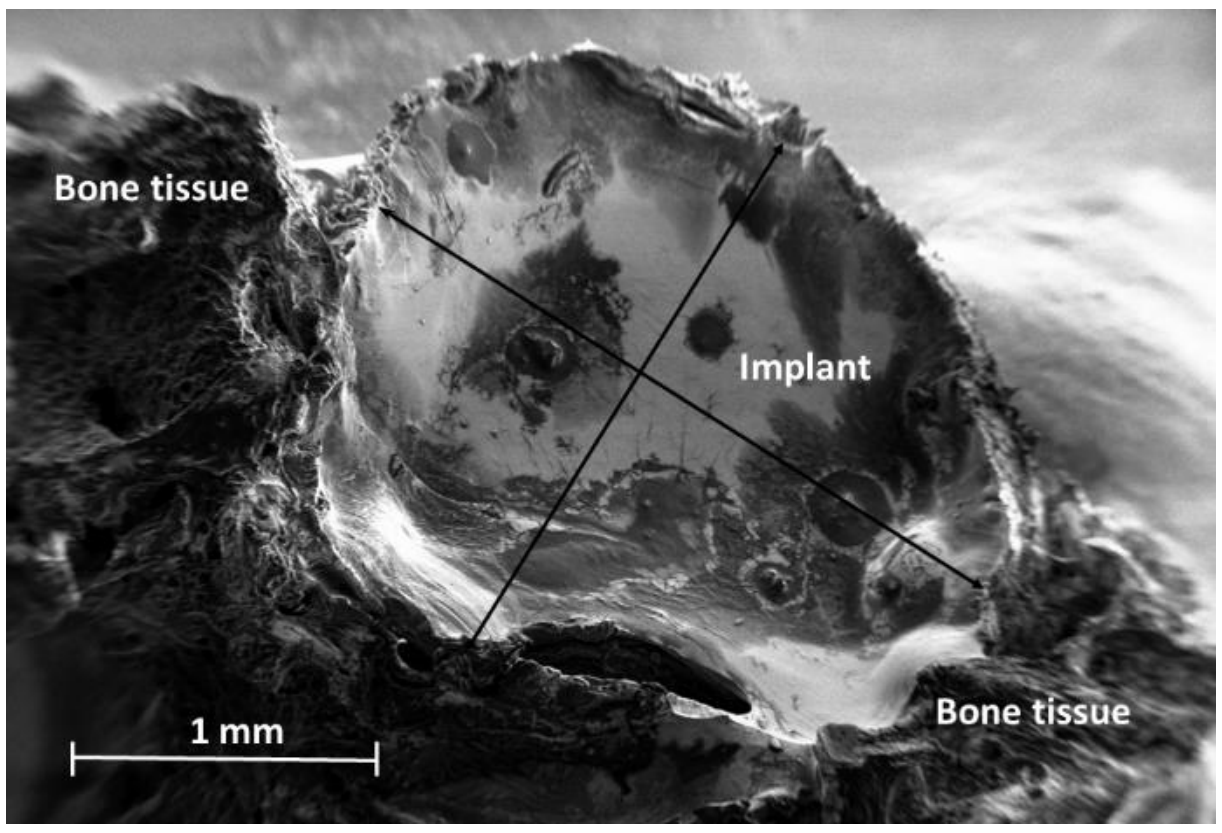

Figure 15. SEM microphotograph of the fractured surface of ionic implant 4 together with the surrounding bone tissues. Magnification ×70.

A portion of bone tissue separated during the preparation of the fractured surface, and the implant edge without bone tissue can be seen in the upper part of Figure 15.

At higher magnification, two bone beams are observed at the interface between the hydrogel and bone tissue, which penetrate into the hydrogel (Fig. 16).

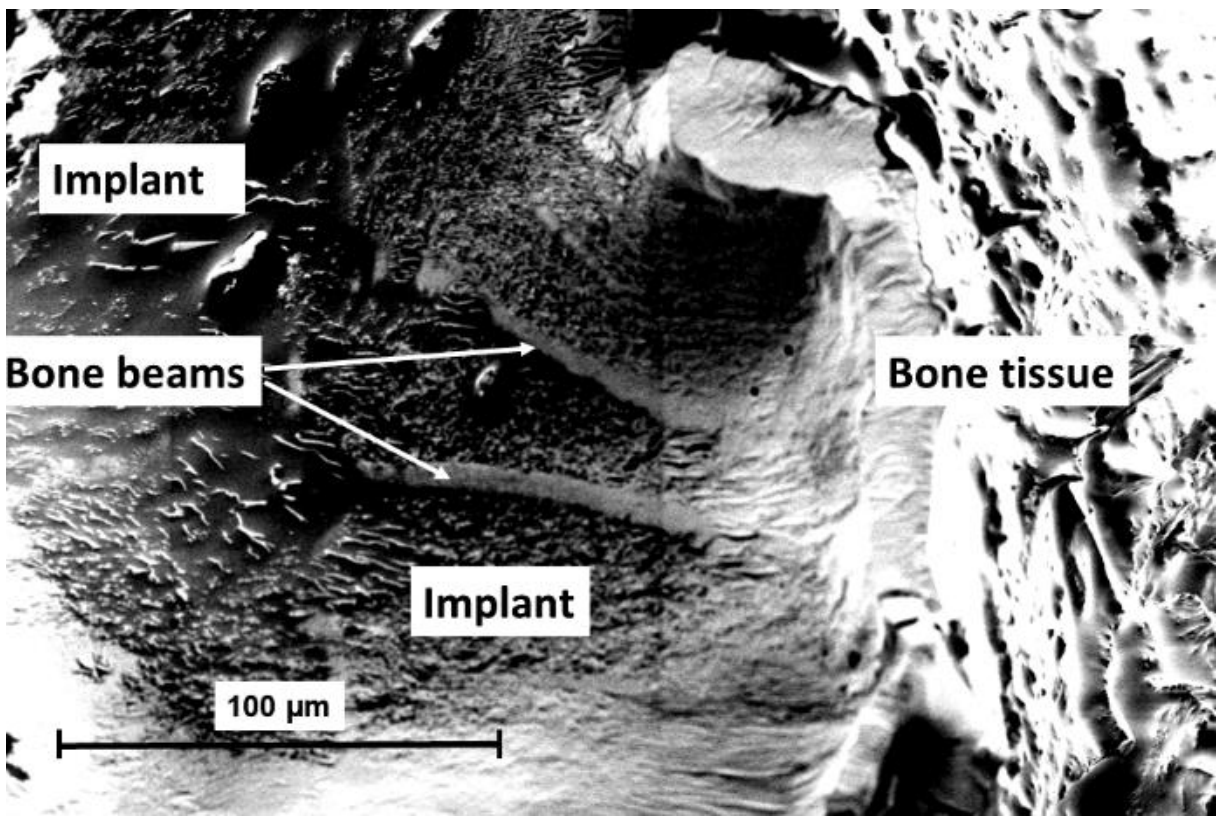

Figure 16. SEM microphotograph of the fractured surface of ionic implant 4 together with the surrounding bone tissues (right part of the image). Magnification ×1000.

The SEM image of another section of the bone-implant interface (Fig. 17) also shows bone beam-like structures marked with arrows.

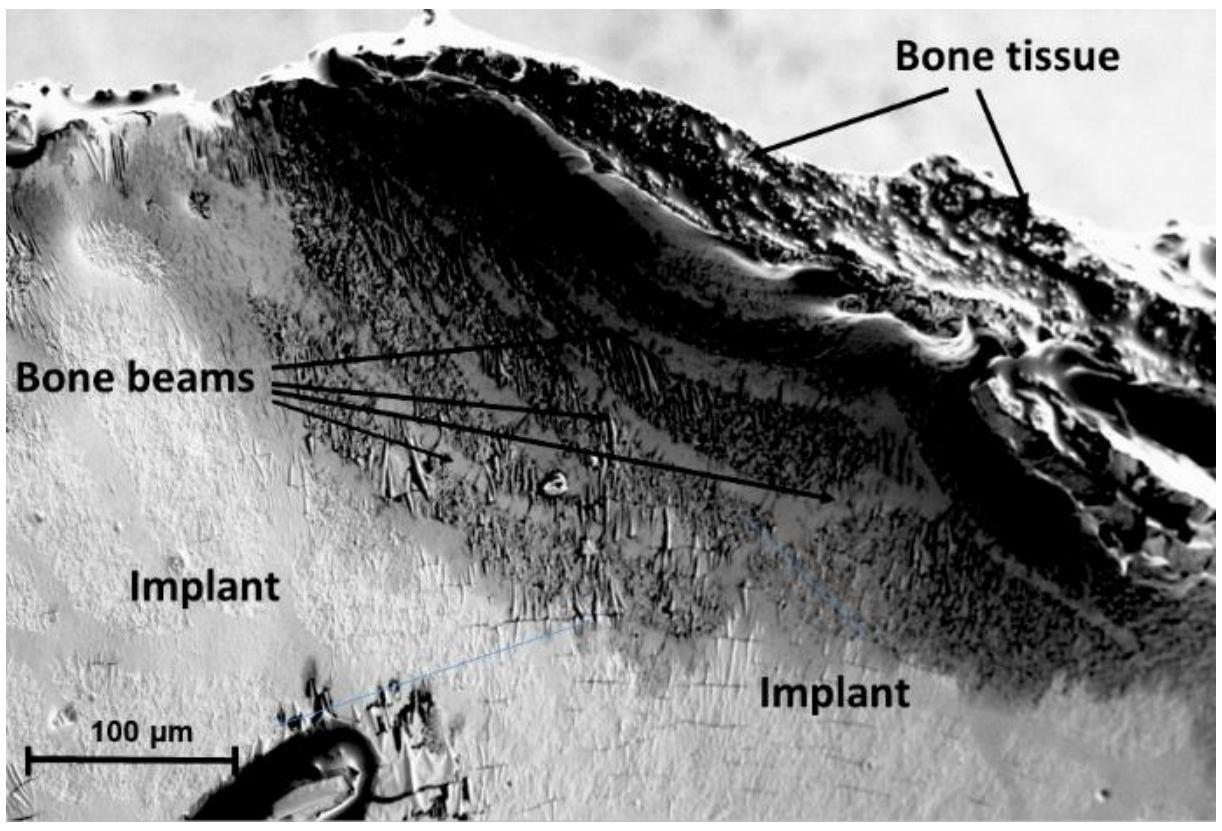

Figure 17. SEM microphotograph of the fractured surface of ionic implant 4 together with the surrounding bone tissues (upper part of the image). Magnification ×500.

All these results demonstrate a very strong integration of the implant with the bone tissue, which actually "grows" into the hydrogel during the *in vivo* experiments (in this case, the duration of experiment was 90 days). Indeed, the bone tissues adhere very tightly to the hydrogel implant (Figs. 15-17) and in some areas the tissue fragments penetrate inside the implant (Figs. 16, 17).

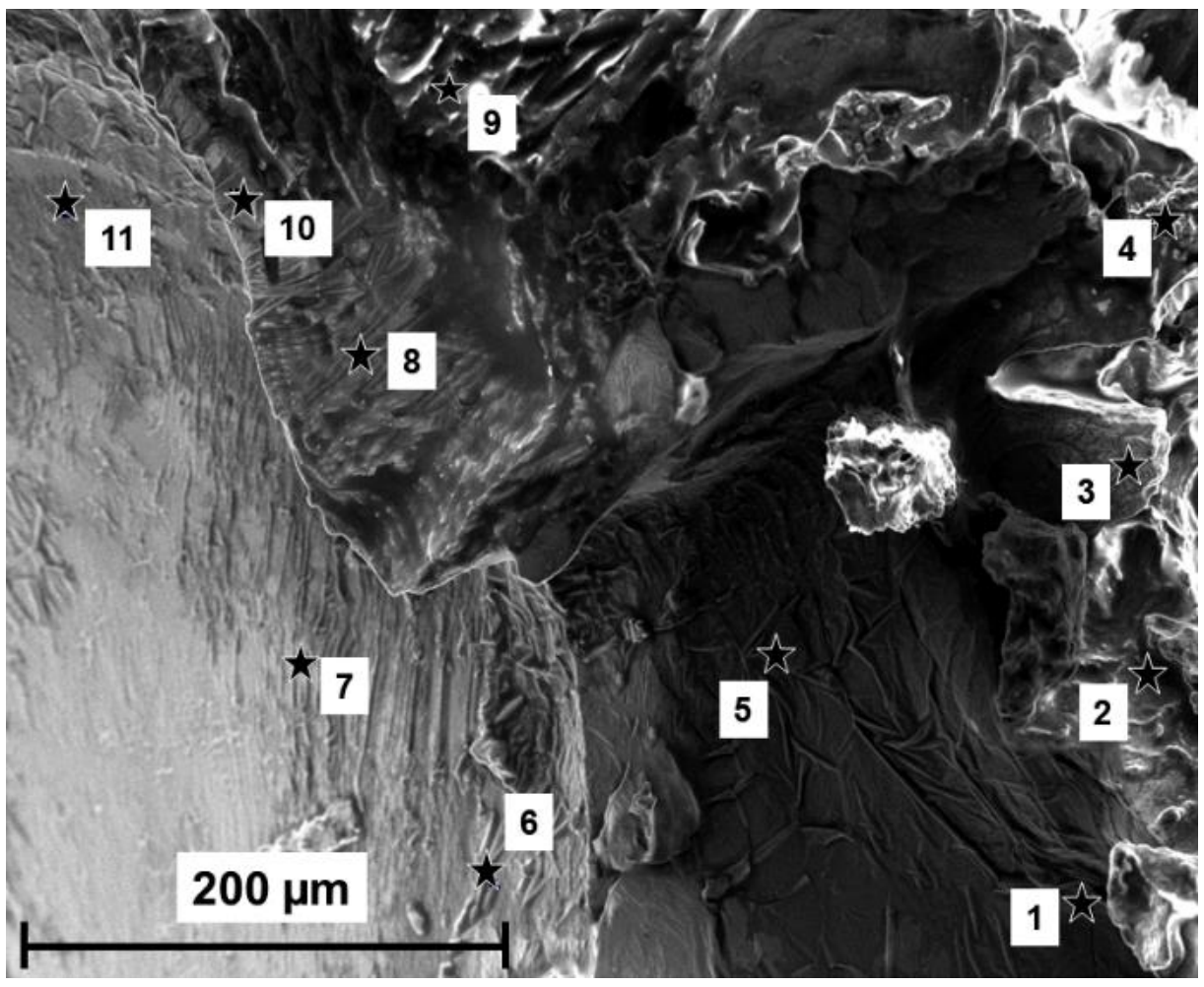

Figure 18. SEM microphotograph of the fractured surface of the subchondral area of implant 4 (near the lateral surface of the sample) extracted from the rabbit body after 120 days of the *in vivo* experiment. Loaded area of cartilage. Magnification ×500.

The elemental composition of implant 4 (Fig. 18, Table 7) is close to that of implant 2. Only in scanning points 3 and 6, calcium and phosphorus are present simultaneously in significant amounts: 4.02 and 2.10 wt.% (point 3); 3.36 and 1.78 wt.% (point 6) of calcium and phosphorus, respectively. Moreover, while point 3 is located near the edge of the implant, point 6 is located approximately 300 μm to the left of the edge. That is, in this case there is no regularity in the distribution of calcium and phosphorus depending on the distance from the implant edge, which was also observed for ionic implant 2. The Ca/P ratio is close to 2 only for the two scanning points discussed above (3 and 6); in other cases, this value varies from 1.25 to 8.75. Thus, along with hydroxyapatite, other calcium phosphates may be present in this implant.

Table 7. Chemical composition of the fractured surface of implant 4 at the scanning points indicated in Fig. 18.

| Spectrum No | Concentrations of elements, wt.%. | | | | | | | |
|---|---|---|---|---|---|---|---|---|
| | C | N | O | Na | P | S | Ca | Ca/P |
| | Initial implant | | | | | | | |
| | 65.73 | 12.39 | 21.17 | 0.71 | 0.00 | 0.00 | 0.00 | - |
| | Implant after *in vivo* experiment | | | | | | | |
| 1 | 64.58 | 11.77 | 22.82 | 0.41 | 0.00 | 0.00 | 0.42 | - |
| 2 | 64.61 | 9.56 | 21.19 | 0.00 | 0.44 | 0.35 | 3.85 | 8.75 |
| 3 | 60.08 | 9.63 | 23.82 | 0.35 | 2.10 | 0.00 | 4.02 | 1.91 |
| 4 | 66.35 | 10.44 | 22.67 | 0.20 | 0.00 | 0.15 | 0.19 | - |
| 5 | 59.84 | 12.54 | 26.81 | 0.34 | 0.00 | 0.08 | 0.39 | - |
| 6 | 63.97 | 8.85 | 21.84 | 0.15 | 1.78 | 0.05 | 3.36 | 1.89 |
| 7 | 63.01 | 10.36 | 26.23 | 0.04 | 0.12 | 0.09 | 0.15 | 1.25 |
| 8 | 65.75 | 10.00 | 22.67 | 0.03 | 0.55 | 0.03 | 0.97 | 1.76 |
| 9 | 63.03 | 13.47 | 22.62 | 0.38 | 0.00 | 0.10 | 0.40 | - |
| 10 | 63.35 | 11.77 | 22.05 | 0.39 | 0.00 | 0.00 | 2.44 | - |
| 11 | 64.84 | 12.54 | 21.81 | 0.30 | 0.00 | 0.08 | 0.43 | - |

It must be also kept in mind that in ionic implants calcium ions can bind to carboxyl groups as counterions, replacing sodium ions. Considering the valence of calcium ions, the formation of bridging ionic bonds between two neighboring carboxyl groups is possible; ionic crosslinking between polymer chains can also occur. The EDX data give the total content of calcium, which can exist either in the form of phosphates or as counterions of the negatively charged carboxyl groups.

In addition, a more thorough analysis of the content of elements in ionic implant 4 was carried out; their average content in rather large areas (highlighted in Fig. 19 by rectangles) was determined (spectra 1 – 3, Table 8).

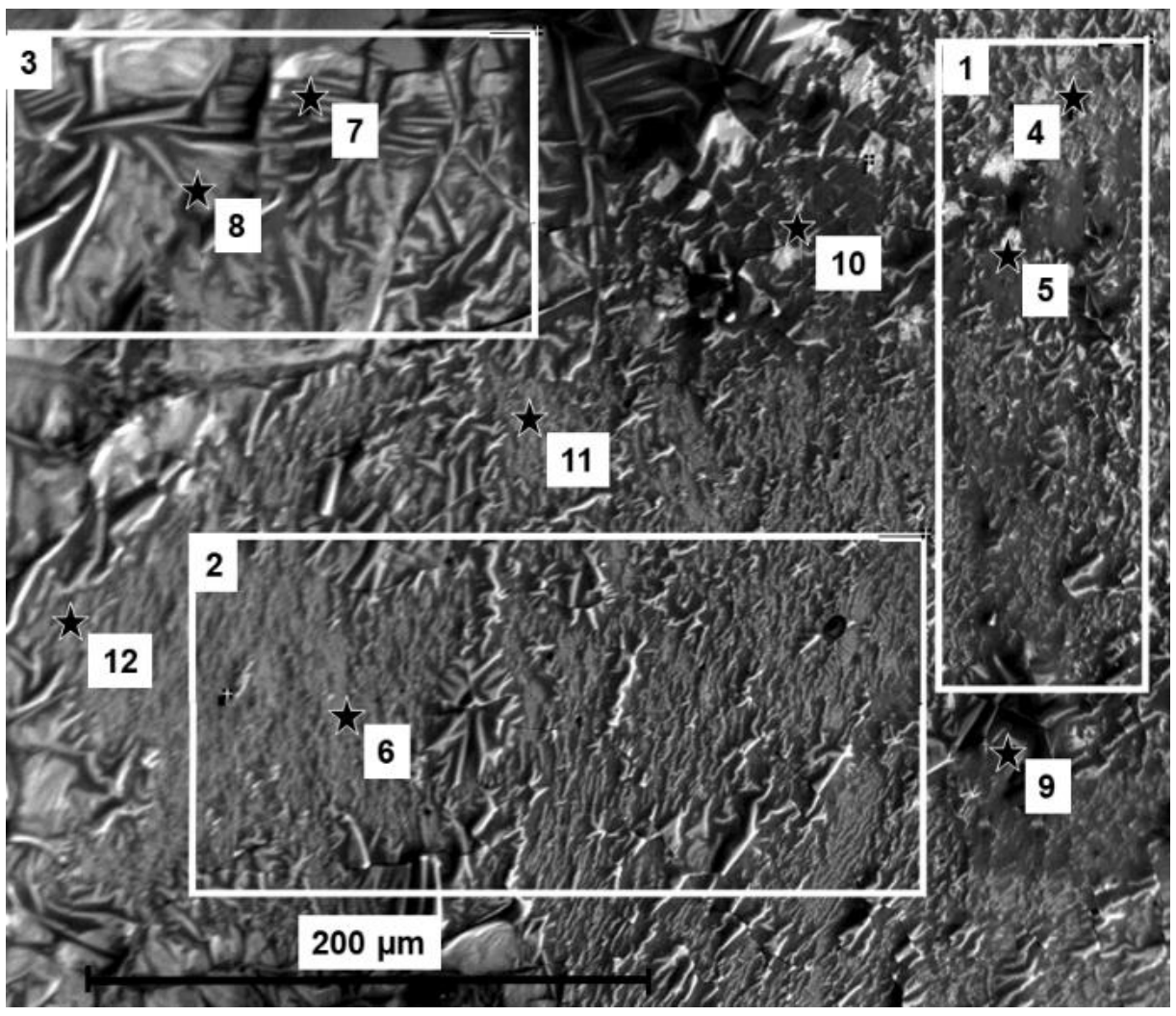

Figure 19. SEM microphotograph of the fractured surface of the subchondral area of implant 4 (near the lateral surface of the sample) extracted from the rabbit body after 120 days of the *in vivo* experiment. Loaded area of cartilage. Magnification ×500.

Table 8. Chemical composition of the fractured surface of implant 4 at the scanning points indicated in Fig. 19.

| Spectrum No | Concentrations of elements, wt.%. | | | | | | |
|---|---|---|---|---|---|---|---|
| | C | N | O | Na | P | K | Ca |
| Initial implant | | | | | | | |
| | 65.73 | 12.39 | 21.17 | 0.71 | 0.00 | 0.00 | 0.00 |
| Implant after *in vivo* experiment | | | | | | | |
| 1 | 66.48 | 12.19 | 20.41 | 0.20 | 0.00 | 0.24 | 0.48 |
| 2 | 65.85 | 11.70 | 20.65 | 0.21 | 0.00 | 0.34 | 1.25 |
| 3 | 67.49 | 10.06 | 20.65 | 0.00 | 0.00 | 0.00 | 1.80 |
| 4 | 62.41 | 10.40 | 23.29 | 0.53 | 0.00 | 0.86 | 2.51 |
| 5 | 62.06 | 10.82 | 23.71 | 0.66 | 0.00 | 0.70 | 2.05 |
| 6 | 65.90 | 10.77 | 20.42 | 0.21 | 0.40 | 0.00 | 2.30 |
| 7 | 65.97 | 10.00 | 21.31 | 0.00 | 0.79 | 0.00 | 1.93 |
| 8 | 63.65 | 9.05 | 22.35 | 0.00 | 0.77 | 0.68 | 3.50 |
| 9 | 64.45 | 13.38 | 21.03 | 0.41 | 0.00 | 0.29 | 0.44 |
| 10 | 63.54 | 9.81 | 22.52 | 0.53 | 0.00 | 1.04 | 2.56 |
| 11 | 63.37 | 12.80 | 20.43 | 0.67 | 0.00 | 0.79 | 1.94 |
| 12 | 65.44 | 12.38 | 21.04 | 0.41 | 0.00 | 0.29 | 0.44 |

As seen from the data presented in Table 8, the calcium content in these highlighted areas is low and varies from 0.48 to 1.8 wt.%, while the phosphorus content is equal to zero. At the same time, phosphorus is found at individual scanning points in amounts up to 0.79 wt.% (spectra 6-8). However, the presence of phosphorus at individual points does not contribute noticeably to the average concentration of this element, which is close to zero.

Thus, a fundamental difference in the degrees of mineralization of ionic and non-ionic types of composite hydrogels is shown. In ionic hydrogels, very low amounts of calcium phosphates are formed in the entire implant volume, whereas in non-ionic PC-PAAm hydrogels the content of calcium phosphates in the area contacted with the subchondral bone reaches 40 wt.%.

Most likely, this effect is due to the existence of the Donnan equilibrium between the gel containing ionic groups and the surrounding solution. For gels, this effect is discussed in detail in the monograph by Flory [39]. As a result of electrostatic factors, the concentration of ions inside the gel decreases compared to their concentration in the external solution. This decrease is more significant the higher the concentration of ionic groups in the polymer network of the hydrogel. In our case, the concentration of ionic groups in implant 2 is approximately 0.3 mol/l, which is two orders of magnitude higher than the calcium concentration in body fluids, which does not exceed 3 mmol/l [45]. At such ratios between the concentrations of charged groups of the polymer network and ions in the external solution (or medium), the ions of solution will not be able to penetrate into the ionic hydrogel implant, i.e. their concentration inside the hydrogel will be close to zero. Assuming that the cells of the organism penetrate into the boundary layers of the hydrogel and facilitate the formation of calcium phosphates inside the hydrogel, the amounts of calcium ions and phosphates in the ionic hydrogel implants will be insufficient for mineralization.

## Conclusions

The samples of studied hydrogels based on plant cellulose and polyacrylamide were implanted into the area of deep cartilage defects of the rabbit knee joints and functioned effectively as artificial cartilages for 90 – 120 days providing complete preservation of the function of the operated extremity. No signs of migration or disintegration of the tested implants were revealed.

The study of the properties of hydrogel implants extracted from rabbit joints 90 – 120 days after implantation showed that their mechanical characteristics remained practically unchanged. The extracted implants, as well as the initial hydrogels, endured cyclic compression loading at the amplitude of 50 %. Compression stresses up to 3 – 10 MPa were recorded in these tests, which is close to the data obtained by several authors for natural articular cartilages.

As for ionic hydrogel implants, no intensive mineralization in the area located in the bone was observed 90 days after their implantation into joints (unlike non-ionic implants). Presumably, mineralization of ionic hydrogels is hindered due to the Donnan equilibrium between the gel and the environment of the organism, which results in a sharp decrease in the concentration of free ions inside the hydrogel compared to their concentration in the environment.

Despite the difference in mineralization rates between non-ionic and ionic hydrogels, good integration of the implants with the subchondral bone of the knee joint of laboratory animals was observed in both cases.

The SEM data also revealed a very strong integration of the implants with bone tissues; bone virtually "grew" into the hydrogel during the *in vivo* experiments. It is evident that mineralization of implants does not play a decisive role in their ability to integrate with bone tissues.

The possibility of controlling the rate of hydrogel implant mineralization by introducing ionic carboxylate groups into synthetic polymer chains was shown.

The substantial depression of the rate of mineralization if the ionic hydrogels are used as the implants can provoke the minimization the possibility of mineral phase formation in the cartilage area of implants, which, in turn, can be important for the long-term functioning of implants as substitutes of cartilage tissues.

**Conflict of Interests**: The authors declare no interest conflict.

**Funding**: This work was financed by the Russian Science Foundation (grant number 22-13-00068).

## *Graphical Abstract*

**Properties of promising cartilage implants based on cellulose/polyacrylamide composite hydrogels: results of *in vivo* tests carried out over a period of 90 – 120 days**

Alexander L. Buyanov, Iosif V. Gofman, Svetlana A. Bozhkova, Natalia N. Saprykin, Georgii I. Netyl'ko, Evgenii F. Panarin

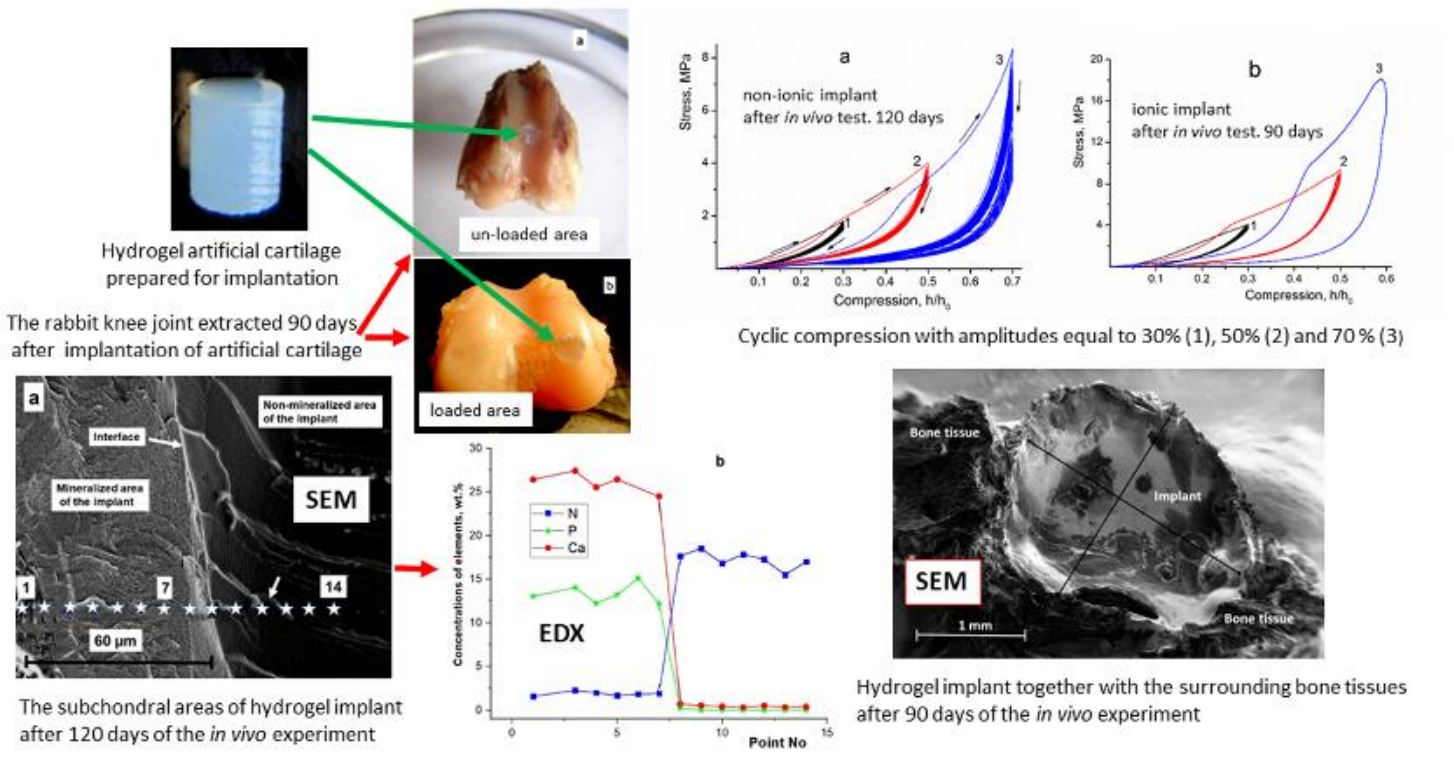